\documentclass[a4paper,12pt]{article}

\pdfoutput=1

\usepackage[a4paper,text={16.8cm,22.8cm}]{geometry}
\usepackage{amsmath,amssymb,bm,graphicx,color}
\usepackage{extarrows}
\usepackage[utf8]{inputenc}
\usepackage{amsthm}
\usepackage{slashed}
\usepackage{tikz}
\usepackage{rotating}
\usepackage{hyperref}
\usepackage{array}
\usepackage{color}
\usepackage{multirow}
\usepackage{bbold}
\usepackage{cite}
\usepackage{pifont}
\usepackage{tabularx}

\allowdisplaybreaks 
\renewcommand{\arraystretch}{1.2}

\newcommand{\xmark}{\ding{55}}

\begin{document}

\begin{titlepage}

\vspace*{-1.0cm}
\begin{flushright} 
LAPTH-033/19\\
DESY 19-091
\end{flushright}

\vspace{1.2cm}
\begin{center}
\Large\bf
\boldmath
Finite Family Groups for Fermionic and Leptoquark Mixing Patterns
\unboldmath
\end{center}
\vspace{0.2cm}
\begin{center}
{\large{Jordan Bernigaud$^a$, Ivo de Medeiros Varzielas$^b$, and Jim Talbert$^c$}}\\
\vspace{1.0cm}
{\sl 
${}^a$\,Univ.\ Grenoble Alpes, Univ.\ Savoie Mont Blanc, CNRS, LAPTh,
9 Chemin de Bellevue,\\ F-74000 Annecy, France\\[0.3cm]
${}^b$\,CFTP, Departamento de F\'{i}sica, Instituto Superior T\'{e}cnico, Universidade de Lisboa, Avenida Rovisco Pais 1, 1049 Lisboa, Portugal\\[0.3cm]
${}^c$\,Theory Group, Deutsches Elektronen-Synchrotron (DESY),\\ Notkestrasse 85, 22607 Hamburg, Germany}\\[0.5cm]
{\bf{E-mail}}:  bernigaud@lapth.cnrs.fr, ivo.de@udo.edu, james.talbert@desy.de
\end{center}

\vspace{0.5cm}
\begin{abstract}
\vspace{0.2cm}
\noindent 
We employ a bottom-up and model-independent technique to search for non-Abelian discrete flavour symmetries capable of predicting viable CKM and PMNS matrices alongside of special patterns of leptoquark couplings.  In particular, we analyze  patterns derived when an ultra-violet flavour theory is assumed to break to global Abelian symmetries in Standard Model fermion masses and new Yukawa-like terms sourced by the leptoquark representation.  The phenomenology of different classes of these `simplified models' can be explored without reference to explicit model-building assumptions, e.g. the nature of flavour symmetry breaking or any additional field content associated to it, and are also capable of explaining hints of lepton non-universality in $\mathcal{R}_{K^{(\star)}}$.  Assuming experimentally interesting CKM and PMNS matrix elements, our algorithm finds an abundance of predictive non-Abelian flavour groups and therefore provides promising directions for future model building in the flavoured leptoquark space, regardless of whether the anomalous $\mathcal{R}_{K^{(\star)}}$ measurements withstand further experimental scrutiny.
\end{abstract}
\vfil

\end{titlepage}

%%%%%%%%%%%%%%%%%%%%%%%%%%%%%%%%%%%%%%%%%%%%%%%%%%%%%%%

\tableofcontents
\noindent \makebox[\linewidth]{\rule{16.8cm}{.4pt}}

%%%%%%%%%%%%%%%%%%%%%%%%%%%%%%%%%%%%%%%%%%%%%%%%%%%%%%%
\section{Introduction}
Flavoured phenomena are amongst the best measured, and least theoretically understood, of the Standard Model (SM) of particle physics.  Accounting for Dirac (Majorana) neutrinos, the SM permits at least 20 (22) free parameters associated to fermionic mass and mixing, and all but one (three) have reliable constraints provided by experiment --- early hints at the leptonic Dirac CP-violating phase exist, albeit with large uncertainties (see e.g. \cite{Esteban:2018azc}).  Furthermore, while all of these unexplained free parameters are associated to Yukawa terms, the strong and electroweak interactions of the SM are otherwise flavour blind; gluons, $W^{\pm}$, and $Z$ gauge bosons couple equally to each fermion species.  The SM's flavour expectations are therefore strikingly different between its scalar and vector interactions, with the former furnishing the so-called \emph{flavour problem} described above, and the latter providing opportunities for precision tests of fermion universality through the decays of heavy mesons.

Intriguingly, recent hints from LHCb \cite{Aaij:2019wad,Aaij:2017vbb} indicate deviations from SM predictions through lepton non-universal (LNU) decays of $B$-mesons, in particular in the ratio observables $\mathcal{R}_{K^{(\star)}}$, 
\begin{align}
\label{eq:RK}
{R}_{K^{(\star)}, [a,b]} &= \frac{\int_{a}^{b} \, dq^{2} \, \left[d\Gamma (B \rightarrow K^{(\star)} \mu^{+}\mu^{-})/ dq^{2} \right]}{\int_{a}^{b} \, dq^{2} \, \left[d\Gamma (B \rightarrow K^{(\star)} e^{+}e^{-})/ dq^{2} \right]} \, .
\end{align}
Here $q^2$ is the invariant mass of the dilepton final state, and $\left[a, b\right]$ represent bin boundaries in GeV$^2$. Experimentally, \eqref{eq:RK} is measured as a double ratio with respect to the resonant high-statistics $J/\Psi$ channel for dilepton production, in order to cancel uncertainties in the measurement efficiencies of the signal modes, and is further shown to only probe LNU in flavour changing neutral current (FCNC) decays
 by testing explicit universality in the  $J/\Psi$ production channels, which are observed to be consistent with the SM \cite{Albrecht:2018frt}.  
 Coupling this robust experimental strategy with rather precise predictions in the SM, where scale and other theory uncertainties for the individual decay channels cancel in the ratio \cite{Hiller:2003js}, it is broadly agreed that one can safely regard \eqref{eq:RK} as clean tests of LNU.  Since LHCb results for both $\mathcal{R}_{K}$ and $\mathcal{R}_{K^{\star}}$ deviate individually between 2-3 $\sigma$ from the SM expectation 
 \cite{Hiller:2003js,Bordone:2016gaq} --- cf. Table \ref{tab:experiment}\footnote{\label{MoriondFootnote}While the $\mathcal{R}_{K}$  results from \cite{Aaij:2019wad} remain at a $\sim 2.5\sigma$ tension with the SM if both Run 1 and Run 2 data sets are included, the Run 2 data appears consistent with unity when analyzed alone.} --- it is then worth considering the sorts of new physics that can generate these early hints of LNU.

%------------------------------------------------------------------------------------------
\begin{table} [t]
\centering
{\renewcommand{\arraystretch}{1.5}
\begin{tabular}{|c||c|c|c|}
\hline
Ratio & Bin (GeV$^2$) & Data & Experimental Reference \\
\hline
\hline
${R}_{K}$ & [1.1, 6] & $0.846^{+0.060+0.016}_{-0.054-0.014}$ & \text{LHCb}\,\,\cite{Aaij:2019wad} \\
\hline
 \multirow{2}{*}{${R}_{K^\star}$} & [1.1, 6.0] & $0.685^{+0.113}_{-0.069} \pm 0.047$ & \text{LHCb}\,\,\cite{Aaij:2017vbb} \\
\cline{2-4}
& [0.045, 1.1] &  $0.66^{+0.11}_{-0.07} \pm 0.03$ & \text{LHCb}\,\,\cite{Aaij:2017vbb} \\
\hline
\end{tabular}}
\caption{${R}_{K^{(\star)}}$ as measured by the LHCb collaboration.  Also see footnote \ref{MoriondFootnote}.
}
\label{tab:experiment}
\end{table}
%------------------------------------------------------------------------------------------

Several theory papers have addressed the anomalous data in Table \ref{tab:experiment}, including model-independent fits to the operators of low-energy effective field theory (EFT) \cite{Hiller:2014yaa,Capdevila:2017bsm,Altmannshofer:2017yso,DAmico:2017mtc,Hiller:2017bzc,Ciuchini:2017mik,Alok:2017sui,Ciuchini:2019usw} as well as concrete beyond-the-Standard Model (BSM) constructions employing composite- or multi-Higgs, leptoquark, or $Z^{\prime}$ fields (to name a few) \cite{deMedeirosVarzielas:2019lgb,Gripaios:2014tna,Varzielas:2015iva,Gripaios:2015gra,Bauer:2015knc,Arnan:2016cpy,Hiller:2016kry,Crivellin:2016ejn,Crivellin:2017zlb,Alonso:2017bff,Bonilla:2017lsq,King:2017anf,Aloni:2017ixa,Assad:2017iib,Calibbi:2017qbu,deMedeirosVarzielas:2018bcy,Grinstein:2018fgb, Fornal:2018dqn,Allanach:2018odd,deMedeirosVarzielas:2019okf,Allanach:2019mfl,Bordone:2017bld,Bordone:2018nbg, Becirevic:2018afm, Angelescu:2018tyl, Cornella:2019hct,Datta:2019tuj,Blanke:2018sro}.  In what follows we 
explore scenarios where the SM's flavour problem is addressed alongside of $\mathcal{R}_{K^{(*)}}$,\footnote{Note however that the formalism we develop is generic, and can be applied to other Lagrangians addressing different combinations of experimental signals.} and we will do so by incorporating one of the following leptoquark representations into the SM Lagrangian:
\begin{equation}
\label{eq:leptoreps}
\Delta_{3} \sim \left(\bar{3}, 3, 1/3\right),\,\,\,\,\,\Delta_{1}^{\mu} \sim \left(3, 1, 2/3\right),\,\,\,\,\,\Delta_{3}^{\mu} \sim \left(3, 3, 2/3\right),
\end{equation}
where the charges given are those of the SM gauge group defined by $\mathcal{G}_{SM} \equiv SU(3)_C \times SU(2)_L \times U(1)_Y$.  We will respectively refer to the states in \eqref{eq:leptoreps} as the scalar triplet, vector singlet, and vector triplet, and all can account for $\mathcal{R}_{K^{(*)}} < 1$ \cite{Hiller:2017bzc}. Moreover, they can be motivated by extended symmetry frameworks like unification theories or new gauge interactions \cite{Perez:2013osa, Biggio:2016wyy}. When added to the field content of the SM they source the following new  $\mathcal{G}_{SM}$-invariant terms in the Lagrangian:
\begin{align}
\nonumber
\Delta_{3}&: \,\,\,\,\,&&\mathcal{L} \supset y_{3, ij}^{LL} \bar{Q}_{L}^{C\,i,a} \epsilon^{ab} (\tau^{k} \Delta_{3}^{k})^{bc} L_{L}^{j,c} + z_{3,ij}^{LL} \bar{Q}_{L}^{C\,i,a}\epsilon^{ab}((\tau^{k}\Delta_{3}^{k})^{\dagger})^{bc}Q_{L}^{j,c} + \text{h.c.} \\
\nonumber
\Delta_{1}^{\mu}&: \,\,\,\,\,&&\mathcal{L} \supset x_{1,ij}^{LL} \bar{Q}_{L}^{i,a} \gamma^{\mu} \Delta_{1,\mu} L_{L}^{j,a} + x_{1,ij}^{RR} \bar{d}^{i}_{R} \gamma^{\mu} \Delta_{1,\mu} e_{R}^{j} + x_{1,ij}^{\overline{RR}} \bar{u}_{R}^{i} \gamma^{\mu} \Delta_{1,\mu} \nu_{R}^{j} + \text{h.c.}\\
\label{eq:LLyukSU2}
\Delta_{3}^{\mu}&: \,\,\,\,\,&&\mathcal{L}  \supset  x_{3,ij}^{LL} \bar{Q}_{L}^{i,a} \gamma^{\mu} \left(\tau^{k} \Delta_{3,\mu}^{k}\right)^{ab} L_{L}^{j,b}  + \text{h.c.} 
\end{align}
where $\lbrace a,b \rbrace$ are SU(2) indices, $\lbrace i,j \rbrace$ are flavour indices, and $k = 1,2,3$ for the Pauli matrices.  As can be seen, the scalar triplet generates a diquark operator that can source proton decay, and the vector singlet introduces new physical interactions between right-handed (RH) fields --- see \cite{Dorsner:2016wpm} for a thorough review of the physics of leptoquarks.

Critically, the coefficients in \eqref{eq:LLyukSU2} are 3 $\times$ 3 complex matrices in flavour space, just like the Yukawa couplings of the SM.  Particular textures in (e.g.) $x_{1,3}^{LL}$ or $y_{3}^{LL}$ will then generate different phenomenology \cite{Hiller:2018wbv,Diaz:2017lit,Schmaltz:2018nls,Baker:2019sli}, and so only special patterns for these couplings are capable of explaining $\mathcal{R}_{K^{(*)}} < 1$ (or any other observable sensitive to their inclusion).  Predictions in such models therefore require that one either $1)$ \emph{assume} a particular form for $x_{1,3}^{LL}$, $y_{3}^{LL}$ or $2)$ \emph{structure} them within an extended theoretical framework, perhaps including a flavour symmetry $\mathcal{G_{F}}$.  Only the latter allows one to simultaneously address the observed scalar and vector LNU, and to that end multiple collaborations have attempted specific `flavourings' of the SM and its $\mathcal{R}_{K^{(\star)}}$-inspired leptoquark extensions (see e.g.  \cite{Fornal:2018dqn,Varzielas:2015iva,Hiller:2016kry,Crivellin:2016ejn,deMedeirosVarzielas:2018bcy,Grinstein:2018fgb,deMedeirosVarzielas:2019okf,Bordone:2017bld,Bordone:2018nbg,Cornella:2019hct,deMedeirosVarzielas:2019lgb,Datta:2019tuj}).  Our goal is to instead determine what sorts of $\mathcal{G}_{\mathcal{F}}$ can generate successful patterns of CKM, PMNS, and leptoquark mixing matrices (associated to $x_{1,3}^{LL}$, $y_{3}^{LL}$) in a model-\emph{independent} fashion.

Although we want to determine viable $\mathcal{G_{F}}$ without committing to specific model-building assumptions, e.g. the dynamics of flavour symmetry breaking, we will focus on a particular class of  $\mathcal{G_{F}}$:  non-Abelian discrete symmetries (NADS), which are well-motivated by both infrared (IR) and ultra-violet (UV) physics.  
Furthermore, we will study NADS in the context of the residual flavour symmetry (RFS) mechanism, where one assumes that $\mathcal{G_{F}}$ breaks to global Abelian flavour symmetries $\mathcal{G}_{a}$ ($a \in \lbrace u,d,l, \nu \rbrace$) in some or all of the SM mass terms and (now) also the leptoquark-sourced terms in \eqref{eq:LLyukSU2}.  The residual $\mathcal{G}_{a}$ then control the shapes of the relevant Yukawa-like couplings in the IR, and the specific forms of the generators that action them can be used to `reconstruct' the parent $\mathcal{G_{F}}$.  The RFS framework generalizes the symmetry-breaking patterns of entire classes of popular flavour models, and as a result has become a useful tool for studying flavour both analytically and numerically within the SM \cite{Lam:2007qc, Ge:2011ih, Ge:2011qn, deAdelhartToorop:2011re, Hernandez:2012ra, Lam:2012ga, Holthausen:2012wt, Holthausen:2013vba, King:2013vna, Lavoura:2014kwa,Fonseca:2014koa,Hu:2014kca, Joshipura:2014pqa, Joshipura:2014qaa, Talbert:2014bda, Yao:2015dwa, King:2016pgv, Varzielas:2016zuo,Yao:2016zev,Lu:2016jit,Li:2017abz,Lu:2018oxc,Hagedorn:2018gpw,Lu:2019gqp} --- reviews can be found in \cite{King:2013eh,Altarelli:2010gt,Grimus:2011fk}.  In fact, two of us recently used RFS to define a novel set of `Simplified Models of Flavourful Leptoquarks' \cite{deMedeirosVarzielas:2019lgb}, where (highly-restrictive) consequences were derived when the \emph{same} RFS representations are assumed to act in SM and leptoquark terms.  However,  in \cite{deMedeirosVarzielas:2019lgb} we did not use the generators of $\mathcal{G}_{a}$ to reconstruct viable $\mathcal{G_{F}}$.  Here we perform this closure using a bottom-up and automated technique developed in \cite{Talbert:2014bda,Varzielas:2016zuo}, both for the symmetry breaking described in \cite{deMedeirosVarzielas:2019lgb} and for a highly natural relaxation of it, i.e. where the leptoquark couplings are driven by only one residual symmetry relation (these models are highly natural in a simple flavon EFT enhancement).  The method employs scripts written with the computational finite algebra package {\tt{GAP}} \cite{Sch97,GAP4}, and we will use them to scan over NADS capable of sourcing interesting phenomenology.  

Importantly, our approach is applicable to any flavoured leptoquark scenario, and therefore will remain relevant regardless of the experimental status of $\mathcal{R}_{K^{(\star)}}$.  In particular, the method developed here allows one to structure the leptoquark patterns while addressing the SM fermionic mixing by identifying different suitable discrete non-Abelian flavour symmetries. This may be used for instance if other deviations from the SM flavour predictions are observed, like for instance $\mathcal{R}_{D^{(\star)}}$. Moreover, our approach is the first use of a flavour symmetry scan for BSM application, which could pave the way for further studies on different models.

The paper develops as follows:  In Section \ref{sec:REC} we review the RFS mechanism, first in the context of the SM alone and then when leptoquarks are included.  We also distinguish two specific symmetry-breaking environments (labeled SE1 and SE2) to scan over, and further derive the `leptoflavour basis' where all relevant physical mixings in the theory can be communicated to our {\tt{GAP}} scripts.  In Section \ref{sec:reconstruction} we review our bottom-up approach for scanning NADS and give details regarding the current BSM leptoquark application.  Finally, we perform the {\tt{GAP}} scans for SE1 and SE2 respectively in Sections \ref{sec:pheno}-\ref{sec:relaxscan}, where additional details relevant to each are presented and a plethora of $\mathcal{G_{F}}$ are discovered.  Closing remarks are provided in Section \ref{sec:conclude}.

%%%%%%%%%%%%%%%%%%%%%%%%%%%%%%%%%%%%%%%%%%%%%%%%%%%%%%%%%%%%
\section{Residual Flavour Symmetries with Leptoquarks}
\label{sec:REC}
Before continuing to study the SM when enhanced by a new leptoquark field, we first review the RFS mechanism in the context of the SM alone  \cite{Lam:2007qc, Ge:2011ih, Ge:2011qn, Hernandez:2012ra,deAdelhartToorop:2011re, Fonseca:2014koa, Hu:2014kca,Lam:2012ga, Holthausen:2012wt, Holthausen:2013vba, King:2013vna, Lavoura:2014kwa, Joshipura:2014pqa, Joshipura:2014qaa, Talbert:2014bda, Yao:2015dwa, King:2016pgv, Varzielas:2016zuo,Yao:2016zev,Lu:2016jit,Li:2017abz,Lu:2018oxc,Hagedorn:2018gpw,Lu:2019gqp}.  As stated above, the core assumption in the RFS framework is that, regardless of the symmetry-breaking mechanism or any dynamics associated to it, a UV flavour symmetry $\mathcal{G_{F}}$ breaks to global Abelian flavour symmetries $\mathcal{G}_{a}$ in some or all of the SM mass terms:
\begin{equation}
\label{eq:GF}
\mathcal{G_{F}}  \rightarrow \begin{cases}
				\mathcal{G_{L}}   \rightarrow \begin{cases} 
										\mathcal{G_{\nu}}
										\\
										\mathcal{G_{\text{l}}}
										\end{cases} \\
				\mathcal{G_{Q}} \rightarrow \begin{cases}
										\mathcal{G_{\text{u}}}
										\\
										\mathcal{G_{\text{d}}}
										\end{cases}
				\end{cases}
\end{equation}
where for illustration we have sketched a symmetry-breaking chain to all four fermion families through two intermediate non-Abelian symmetries $\mathcal{G}_{\mathcal{L},\mathcal{Q}}$ that control only leptons or quarks.  Other breaking patterns are of course also conceivable.
Regardless, the scenario outlined in \eqref{eq:GF} appears quite natural as, after all, the mass terms of SM charged fermions and (if present) Dirac neutrinos already exhibit accidental $U(1)^3$ global symmetries associated to independent rephasings of each generation. If neutrinos are instead Majorana fields they respect an accidental $\mathbb{Z}_{2} \times \mathbb{Z}_{2}$ Klein symmetry.  To see this explicitly we write down the SM Yukawa sector after EWSB, in the fermion mass basis:
\begin{align}
\label{eq:SMfermionmass}
\mathcal{L}^{SM}_{mass} \,\,\, \supset \,\,\,&\frac{1}{2} \bar{\nu}^{c}_{L}\, m_{\nu}\, \nu_{L} + \bar{E}_{R}\, m_{l}\, l_{L} + \bar{d}_{R}\, m_{d}\, d_{L} + \bar{u}_{R}\, m_{u} \,u_{L} +\, \text{h.c.}
\end{align}
where for now we have included a Majorana neutrino mass term, as may be generated in a Type-I seesaw mechanism \cite{Minkowski:1977sc}, to illustrate our point.  Here $m_{a}$ are all diagonal matrices of mass eigenvalues.  We now observe that \eqref{eq:SMfermionmass} is invariant under the following operations on its fields:
\begin{align}
\nonumber
\nu_{L} &\rightarrow T_{\nu_{i}}\nu_{L}, \,\,\,\,\,&&\text{with} \,\,\,\,\,&&T_{\nu 1} = diag \left(1,-1,-1 \right) &&\text{and}\,\,\,\,\, T_{\nu 2} = diag \left(-1,1,-1 \right),\\
\label{eq:SMRFS}
f &\rightarrow T_{f} f, \,\,\,\,\,&&\text{with} \,\,\,\,\,&&T_{f} = diag \left(e^{i \alpha_{f}},e^{i \beta_{f}},e^{i \gamma_{f}} \right)\,\,\,\,\,&&\text{for}\,\,\,\,\, f \in \lbrace E_{R}, l_{L}, d_{R}, d_{L}, u_{R}, u_{L} \rbrace.
\end{align}
In \eqref{eq:SMRFS} we have simply arranged the action of the aforementioned accidental Abelian symmetries into (reducible) triplet representations whose diagonal elements distinguish different generations.  To clarify, in doing this (and throughout the text) we are always considering that the abstract group element of the parent $\mathcal{G_F}$, such as $T_f$ above, is implicitly represented by a (reducible or irreducible) triplet representation in $\mathcal{G_F}$. This necessarily decomposes into irreducible singlets of the residual symmetries, which are Abelian subgroups of $\mathcal{G_F}$ and would simply correspond to multiplication by the respective phases (e.g. in the Tables of Section \ref{sec:pheno}, \ref{sec:relaxscan}). Clearly $T_{\nu_{1,2}}$ generate the Klein four-group and $T_{f}$ generates the respective $U(1)^{3}$ of Dirac fermions.\footnote{Note that, in a generic flavour symmetry framework, the right-handed (RH) fermions need not transform under the same representation as the left-handed (LH) ones. It is after the flavour symmetry is broken (either to residual subgroups or not) that the mass term requires LH and RH fermions to transform in a related way.}  If one instead wishes to identify a discrete subgroup of $U(1)^3$, as we will below in order to identify NADS, the free phases get quantized as
\begin{equation}
\label{eq:discretizephases}
\lbrace \alpha, \beta, \gamma \rbrace_{f} \overset{!}{=} \frac{2 \pi}{m} \lbrace a, b,  c  \rbrace_{f}
\end{equation}
with $m$ the order of the cycle symmetry $\mathbb{Z}_{m}$ being generated.  Cyclic product subgroups with more than one generator are also possible and potentially interesting.

Critically, in the RFS framework, the symmetries described by \eqref{eq:SMRFS} are no longer accidental --- they represent the infrared (IR) signatures of a complete flavour theory controlled by $\mathcal{G_{F}}$, which commutes with the entire SM (or any BSM completion, e.g. an $SU(5)$ grand unified theory).  For example, $\mathcal{G}_{a}$ may appear when, in some or all SM Yukawa operators, scalar flavon fields break $\mathcal{G_{F}}$ via vacuum expectation values (VEVs) aligned along special directions of flavour space.  Thinking from the top down, these special alignments (and therefore the particular $\mathcal{G}_{a}$ realized) are a consequence of the form of a ($\mathcal{G_{F}} \times \mathcal{G_{(B)SM}}$)-invariant scalar potential.  On the other hand, from a bottom-up perspective, different phase configurations for the RFS generators $T_{a}$, once `chosen,' correspond to different (phenomenologically relevant) configurations of fermion mixing matrices.  

This latter point is best seen in the SM flavour basis, where the charged-current interactions of the SM are diagonal, but its mass matrices are not:
\begin{align}
\label{eq:SMfermionflav}
\mathcal{L}^{SM}_{flav} \,\,\, \supset \,\,\,&\frac{1}{2} \bar{\nu}^{c}_{L}U^{\star}_{\nu} m_{\nu} U^{\dagger}_{\nu} \nu_{L} + \bar{E}_{R}U_{E} m_{l}U_{l}^{\dagger} l_{L} + \bar{d}_{R}U_{D} m_{d} U_{d}^{\dagger} d_{L} + \bar{u}_{R}U_{U}  m_{u} U_{u}^{\dagger} u_{L} +\, \text{h.c.}
\end{align}
The $U$ transformations are 3 $\times$ 3 unitary matrices, and the physical CKM and PMNS mixing matrices of the SM are defined in terms of those acting on the LH fields participating in the charged interactions:
\begin{equation}
\label{eq:CKMandPMNS}
U_{CKM} \equiv U^{\dagger}_{u} \, U_{d}, \,\,\,\,\,\,\,\,\,\,U_{PMNS} \equiv U^{\dagger}_{l} \, U_{\nu}.
\end{equation}
One now observes the following invariance of \eqref{eq:SMfermionflav}:
\begin{align}
\label{eq:SMRFS2}
a &\rightarrow T_{a U}\, a \,\,\,\,\,\text{with} \,\,\,\,\,T_{a U} = U_{a} T_{a} U_{a}^{\dagger}\,,
\end{align}
with $a$ representing all fermions, including neutrinos.  This invariance is interpreted as a symmetry of the mass matrix,
\begin{equation}
m_{a U} = T^{\dagger}_{a U} m_{a U} T_{a U},
\end{equation}
where the Hermitian conjugate `$\dagger$' gets replaced with a transpose `$T$' for Majorana neutrinos. 

One now also sees how the mixing of particle species can be connected directly to the parent group structure.  In \eqref{eq:SMRFS2}, the generators are written explicitly as functions of the physical mixing matrices.  Assuming that our flavour symmetry $\mathcal{G_{F}}$ breaks down to the RFS present in \eqref{eq:GF}, then one can `reconstruct' the $\mathcal{G_{F}}$ as the group generated by $\lbrace T_{\nu iU}, T_{l U}, T_{d U}, T_{u U} \rbrace$ or any allowed combination therein (in the event $\mathcal{G_{F}}$ does not break to all four families).  This bottom-up approach to studying flavour is not merely a mathematical trick.  It describes the symmetry-breaking patterns of entire classes of flavour models,\footnote{These are referred to as `direct' and `semi-direct' models in the taxonomy of \cite{King:2013eh}.  Other `indirect' models, where the accidental symmetries of \eqref{eq:SMRFS} and \eqref{eq:SMRFS2} are not controlled by subgroups of $\mathcal{G_{F}}$, are of course also popular in the flavoured model-building literature --- see \cite{deMedeirosVarzielas:2017sdv} for a successful and recent example.} including the famous Altarelli-Feruglio model of leptonic mass and mixing \cite{Altarelli:2005yx}.  There, $\mathcal{G}_{L} \cong A_{4}$ is broken by flavon fields whose VEVs align themselves in different directions in the charged lepton and neutrino mass terms, leaving residual $\mathbb{Z}_{3,2}$ \footnote{Note that in the original model, one of the $\mathbb{Z}_2$ residual symmetries is accidental and is generated by the absence of a specific flavon representation. In $S_4$ models however, the $\mathbb{Z}_2$ symmetry can be obtained directly as a subgroup of the parent symmetry.}  symmetries (to be identified as $\mathcal{G}_{l,\nu}$) in these respective sectors.  The associated mass-basis generators $T_{l,\nu}$, when rotated through \eqref{eq:SMRFS2} with $U_{a} = U_{TBM}$,\footnote{Note that in \cite{Altarelli:2005yx} the charged lepton mass matrix is already diagonal, so $U_{e} = \mathbb{1}$ and therefore $U_{\nu} = U_{PMNS}$.} the tri-bimaximal mixing matrix \cite{Harrison:2002er} that the model predicts, immediately close the original $A_4$ group!

%%%%%%%%%%%%%%%%%%%%%%%%%%%%%%%%%%%%%%%%%%%%%%%%%%%%%%%%%%%
\subsection{Isospin Decomposition of Leptoquark Couplings}
\label{sec:isospin}
We now wish to extend the above analysis to include the leptoquark representations of \eqref{eq:LLyukSU2}, although for brevity we will typically only show details for the scalar triplet $\Delta_{3}$;  the vector singlet and triplet analyses follow in precisely the same way, and any special caveats will be mentioned when relevant.  Also note that we assume only one new leptoquark in the Lagrangian, which is motivated by our expectation that they are singlets under $\mathcal{G_F}$, the same assignment that Higgs fields typically take in most flavour models, and as they are in the explicit leptoquark models of \cite{Varzielas:2015iva}.  Adding multiple singlets or a multiplet under $\mathcal{G_F}$ would add several intricacies, whereas with a single generation of leptoquarks, we are always dealing with leptoquark mass eigenstates and we avoid replication of coupling matrices between the leptoquark and fermions.

As in \cite{Dorsner:2016wpm}, we define new combinations of the isospin components of $\Delta_{3}$ as
\begin{equation}
\label{eq:decomp}
\Delta_{3}^{4/3} = \left(\Delta_{3}^{1} - i \Delta_{3}^{2} \right)/\sqrt{2}, \,\,\,\,\,\,\,\,\,\,\,\,\,\,
\Delta_{3}^{-2/3} = \left(\Delta_{3}^{1} + i \Delta_{3}^{2} \right)/\sqrt{2}, \,\,\,\,\,\,\,\,\,\,\,\,\,\,
\Delta_{3}^{1/3} = \Delta_{3}^{3}, 
\end{equation}
with exponents denoting electric charges and SU(2) indices on the left- and right-hand sides, respectively.  Contracting SU(2) indices, we can write the scalar triplet Lagrangian in \eqref{eq:LLyukSU2} explicitly in the mass basis of the SM fermions, obtaining 
\begin{align}
\nonumber
\mathcal{L}_{mass}^{LQ}\,\,\,\,\, \supset \,\,\,\,\, &\underbrace{-(U_{d}^{T} y_{3}^{LL} U_{\nu})_{ij}}_{\lambda_{d\nu}} \bar{d}^{C \, i}_{L} \Delta^{1/3}_{3} \nu_{L}^{j} \underbrace{- \sqrt{2} (U_{d}^{T} y_{3}^{LL} U_{l})_{ij}}_{\lambda_{dl}} \bar{d}^{C \, i}_{L} \Delta^{4/3}_{3} l_{L}^{j}  \\
\nonumber
&+ \underbrace{\sqrt{2} (U_{u}^{T} y_{3}^{LL} U_{\nu})_{ij}}_{\lambda_{u\nu}} \bar{u}^{C \, i}_{L} \Delta^{-2/3}_{3} \nu_{L}^{j} \underbrace{-(U_{u}^{T} y_{3}^{LL} U_{l})_{ij}}_{\lambda_{ul}}\bar{u}^{C \, i}_{L} \Delta^{1/3}_{3} l_{L}^{j}  \\
\label{eq:SU2scalar}
&+ \text{h.c.}
\end{align}
where we leave aside the diquark operators, although the residual symmetries can also apply there.\footnote{As discussed in more detail in \cite{deMedeirosVarzielas:2019lgb}, the RFS can also shape the diquark couplings  $z_3^{LL}$ by forcing zeros in the matrix elements controlling first-generation (or indeed all) transitions, and thereby protect against proton decay.  The phase equalities required in the RFS generators $T_{u,d}$ are compatible with those derived from the quark-lepton couplings seen below in Table \ref{tab:finalphases}.}
Here it is clear that the $\lambda_{QL}$ combinations we have defined can all be written in terms of a single coupling,
\begin{equation}
\label{eq:LQrelations}
\lambda_{d \nu} = \frac{1}{\sqrt{2}} \lambda_{dl}\,U_{PMNS}, \,\,\,\,\,\, \lambda_{ul} = \frac{1}{\sqrt{2}} U^{\star}_{CKM}\,\lambda_{dl}, \,\,\,\,\,\, \lambda_{u \nu} = - U^{\star}_{CKM}\,\lambda_{dl}\,U_{PMNS}.
\end{equation}
We have chosen to normalize to $\lambda_{dl}$, the matrix we can constrain via measurements of $\mathcal{R}_{K^{(\star)}}$, and where we have used the definitions of the CKM and PMNS matrices in \eqref{eq:CKMandPMNS}.  The analogous relationships for $\Delta_{3}^{\mu}$ are given by
\begin{align}
\label{eq:LQrelationsV3}
\lambda_{d\nu}^{V_{3}} = - \sqrt{2} \, \lambda_{dl}^{V_{3}} \,U_{PMNS}, \,\,\,\,\,\,\,\,\,\, \lambda_{ul}^{V_{3}} = - \sqrt{2} \, U_{CKM} \,\lambda_{dl}^{V_{3}}, \,\,\,\,\,\,\,\,\,\, \lambda_{u\nu}^{V_{3}} = - U_{CKM} \, \lambda_{dl}^{V_{3}} \, U_{PMNS},
\end{align}
where we have distinguished these from the scalar triplet through the additional `$V_{3}$' label (the conjugation structure of the fields in \eqref{eq:LLyukSU2} yields a slightly different normalization for the $d-l$ coupling: $\lambda_{dl}^{V_{3}} \equiv - (U_{d}^{\dagger} x_{3}^{LL} U_{l})$.).  On the other hand, we only have one such correspondence for the vector singlet, since we do not have RH analogues to the CKM and PMNS matrices:
\begin{equation}
\label{eq:LQrelationsV1}
\lambda_{u\nu}^{V_{1}} =  U_{CKM} \, \lambda_{dl}^{V_{1}} \, U_{PMNS},
\end{equation} 
with the redefined $d-l$ coupling now given by $\lambda_{dl}^{V_{1}} \equiv (U_{d}^{\dagger} x_{1}^{LL} U_{l})$.  As it turns out, the SU(2) relationships in \eqref{eq:LQrelations}-\eqref{eq:LQrelationsV1} are extremely important not only in determining the overall shape of the relevant RFS generators in a chosen basis, but also in restricting the experimentally allowed phases controlling the order of any given generator.

%%%%%%%%%%%%%%%%%%%%%%%%%%%%%%%%%%%%%%%%%%%%%%%%%%%%%%%%%%%%
\subsection{The Fermion Mass Basis}
\label{sec:massbasis}

Including all relevant terms, the full Yukawa sector of our $\Delta_{3}$-enhanced Lagrangian, in the mass basis of the SM fermions, now reads
\begin{align}
\nonumber
\mathcal{L}_{mass} \,\,\, \supset \,\,\,&\frac{1}{2} \bar{\nu}^{c}_{L}\, m_{\nu}\, \nu_{L} + \bar{E}_{R}\, m_{l}\, l_{L} + \bar{d}_{R}\, m_{d}\, d_{L} + \bar{u}_{R}\, m_{u} \,u_{L}   \\
\nonumber
&+\, \bar{d}^{C}_{L} \,\lambda_{dl}\, l_{L} \, \Delta^{4/3}_{3} + \bar{d}^{C}_{L} \,\lambda_{d\nu}\, \nu_{L} \, \Delta^{1/3}_{3} + \bar{u}^{C}_{L} \,\lambda_{ul}\, l_{L} \, \Delta^{1/3}_{3} + \bar{u}^{C}_{L} \,\lambda_{u\nu}\, \nu_{L} \, \Delta^{-2/3}_{3} \\
\label{eq:LQYUKfull}
&+\, \text{h.c.}
\end{align}
with $m_{a}$ diagonal matrices of mass eigenvalues, and the $\lambda_{QL}$ defined as in \eqref{eq:SU2scalar} and \eqref{eq:LQrelations}.  Since we are in the fermion mass basis, the leptoquark Yukawa couplings are generically non-diagonal, with rows and columns identifiable in a generation specific way.  For example, $\lambda_{dl}$ can be written as \cite{Varzielas:2015iva}    
\begin{equation}
\label{eq:genericyuk}
-\sqrt{2}\,\left(U_{d}^{T}\,y_{3}^{LL}\,U_{l} \right) \equiv \lambda_{dl} =
\left(
\begin{array}{ccc}
\lambda_{de} & \lambda_{d\mu} & \lambda_{d\tau}  \\
\lambda_{se}  & \lambda_{s\mu}  & \lambda_{s\tau}     \\
\lambda_{be}  & \lambda_{b\mu}   & \lambda_{b\tau}  
\end{array}
\right) \,.
\end{equation}
Given \eqref{eq:genericyuk}, the starting assumption of our analysis is that
\begin{equation}
\label{eq:symmetryexists}
\exists \,\, \lbrace Q, L \rbrace,\,\,\,\, T^{(T, \dagger)}_{Q}\, \lambda_{QL}\,T_{L} \overset{!}{=} \lambda_{QL},
\end{equation}
where $T_Q$ is transposed `$T$' (daggered `$\dagger$') when considering scalar (vector) leptoquark(s).  That is, we assume that residual symmetries also constrain the matrix elements of at least one leptoquark coupling, and of course in what follows we will always include $\lambda_{dl}$, so that we have theoretical control over $\mathcal{R}_{K^{(\star)}}$.  \eqref{eq:symmetryexists} further implies that the \emph{same} generator representations $T_{a}$ acting on fermion fields in their respective SM mass terms also action the RFS in (at least one of) the new leptoquark couplings.  This assumption is of course not required from the model-building perspective, however it is highly plausible.  So, while building explicit models that realize \eqref{eq:symmetryexists} is beyond the scope (and in fact antithetical to the purpose) of this paper, we will briefly mention possible explanations for its origins below, where  we consider two interesting cases of \eqref{eq:symmetryexists} that have also already been explored in the literature, either directly or indirectly.  Namely, we study \eqref{eq:symmetryexists} in the following `symmetry environments:'

\begin{enumerate}
 \item \label{item1} {\bf{\underline{Symmetry Environment 1 (SE1) --- Fully-Reduced Matrices}:}}  The same RFS hold in all four SM mass terms \emph{and} all four SU(2) related leptoquark couplings.  This scenario corresponds to the `Simplified Models of Flavourful Leptoquarks' presented in detail in \cite{deMedeirosVarzielas:2019lgb}, where it was shown that the arbitrary 3$\times$3 complex matrices of $\lambda_{QL}$ are simplified to matrices with only a single real parametric degree of freedom, as shown in Table \ref{tab:finalphases}.  These `fully-reduced' matrices can be realized, e.g., in effective models where the operators in \eqref{eq:LLyukSU2} are enhanced to include 1) flavon(s) to structure the $\lambda_{QL}$ via their VEVs, and 2) other scalars that can distinguish the members of SU(2) doublets after EWSB (in a way that preserves \eqref{eq:LQrelations}). 
 \item \label{item2} {\bf{\underline{Symmetry Environment 2 (SE2) --- Partially-Reduced Matrices}:}} RFS hold in some or all of the SM mass terms, but only SM down quark and/or charged lepton symmetries are active in the leptoquark sector, controlling the shape of $\lambda_{dl}$.\footnote{We consider then that any other symmetry present is understood as accidental, i.e. not controlled by an explicit subgroup of $\mathcal{G_{F}}$. This scenario is again analogous to the Altarelli-Feruglio model \cite{Altarelli:2005yx}, where the neutrino mass matrix predicted is invariant under a $\mu-\tau$ operator generating a $\mathbb{Z}_{2}$ symmetry that is \emph{not} a subgroup of $A_4$.}  Symmetries are respected by $\lambda_{d\nu,ul,u\nu}$ because they are inherited from $\lambda_{dl}$ via SU(2) relations.
  These represent relaxed versions of the simplified models of \cite{deMedeirosVarzielas:2019lgb}, and generalize the complete models written down in \cite{Varzielas:2015iva}, which are realized by inserting a single flavon in the terms in \eqref{eq:LLyukSU2}, yielding effective operators of mass dimension five.  Hence they do not require additional non-trivial SU(2) scalars, and in this sense may be more minimal than models constructed in SE1.  However, as their name suggests, the resulting $\lambda_{QL}$ have more parametric degrees of freedom --- they are only `partially reduced.'
 \end{enumerate}
In both SE1 and SE2, the $T$ generators are again represented by diagonal matrices with three phases, such that an equality of the following form appears (e.g.) for $\lambda_{dl}$ \cite{deMedeirosVarzielas:2019lgb}:
 \begin{equation}
\label{eq:LQoverconstrain}
    \left(
\begin{array}{ccc}
e^{i(\alpha_{d}+\alpha_{l})}\,\lambda_{de} & e^{i(\alpha_{d}+\beta_{l})}\, \lambda_{d\mu} &  e^{i(\alpha_{d}+\gamma_{l})}\,\lambda_{d\tau}  \\
e^{i(\beta_{d}+\alpha_{l})}\,\lambda_{se}  & e^{i(\beta_{d}+\beta_{l})}\, \lambda_{s\mu}  & e^{i(\beta_{d}+\gamma_{l})}\, \lambda_{s\tau}     \\
e^{i(\gamma_{d}+\alpha_{l})}\,\lambda_{be}  & e^{i(\gamma_{d}+\beta_{l})}\,\lambda_{b\mu}   &  e^{i(\gamma_{d}+\gamma_{l})}\,\lambda_{b\tau}  
\end{array}
\right)
  \overset{!}{=}
  \left(
\begin{array}{ccc}
\lambda_{de} & \lambda_{d\mu} & \lambda_{d\tau}  \\
\lambda_{se}  & \lambda_{s\mu}  & \lambda_{s\tau}     \\
\lambda_{be}  & \lambda_{b\mu}   & \lambda_{b\tau}  
\end{array}
\right) .
\end{equation}
In the event that only quark or lepton symmetries are active in SE2, then only the phases associated to $T_{d}$ or $T_{l}$ are non-zero in \eqref{eq:LQoverconstrain}, respectively.  Importantly, the solutions to \eqref{eq:LQoverconstrain} that are LNU (following the implications of $\mathcal{R}_{K^{(\star)}}$) and which distinguish multiple generations in each family, as would be expected for a family symmetry, are few in number.  

%---------------------------------------------------------------------------------------------------------------------------------------
\begin{table} [htp]
\centering
\small
{\renewcommand{\arraystretch}{1.5}
\begin{tabular}{|c||c|c|c|}
\hline
 \multicolumn{1}{|c||}{$\lambda_{QL}$} & \multicolumn{2}{c|}{Phase Equalities}  & $\lambda_{dl}$ \\
\hline
\hline
\multirow{3}{*}{$\lambda_{QL}^{e3A}$} & $\Delta_{3}$ &$\lbrace$ $\beta_{d}$, $\gamma_{d}$, $-\alpha_{\nu}$, $-\beta_{\nu}$, $-\alpha_{l}$, $\beta_{u}$, $\gamma_{u}$ $\rbrace$ & \multirow{3}{*}{
$ \lambda_{be}
\left(
\begin{array}{ccc}
0  & 0 & 0  \\
-\frac{V_{ub}}{V_{us}} & 0 & 0    \\
1 & 0 & 0 
\end{array}
\right)$
}\\
\cline{2-3}
& $\Delta_{3}^{\mu}$  & $\lbrace$ $\beta_{d}$, $\gamma_{d}$, $\alpha_{\nu}$, $\beta_{\nu}$, $\alpha_{l}$, $\beta_{u}$, $\gamma_{u}$ $\rbrace$ & \\
\cline{2-3}
& $\Delta_{1}^{\mu}$ & $\lbrace$ $\beta_{d}, \gamma_{d}, \alpha_{l}$ $\rbrace$ $\lbrace$ $\alpha_{\nu}, \beta_{\nu}, \beta_{u}, \gamma_{u}$ $\rbrace$ & \\
\cline{1-4}
\multirow{3}{*}{$\lambda_{QL}^{e3B}$} & $\Delta_{3}$ & $\lbrace$ $\beta_{d}$, $\gamma_{d}$, $-\alpha_{\nu}$, $-\beta_{\nu}$, $-\alpha_{l}$, $\alpha_{u}$, $\gamma_{u}$ $\rbrace$ & \multirow{3}{*}{
$\lambda_{be}
\left(
\begin{array}{ccc}
0  & 0 & 0  \\
-\frac{V_{cb}}{V_{cs}} & 0 & 0    \\
1 & 0 & 0 
\end{array}
\right)$
} \\
\cline{2-3}
& $\Delta_{3}^{\mu}$  & $\lbrace$ $\beta_{d}$, $\gamma_{d}$, $\alpha_{\nu}$, $\beta_{\nu}$, $\alpha_{l}$, $\alpha_{u}$, $\gamma_{u}$ $\rbrace$ & \\
\cline{2-3}
& $\Delta_{1}^{\mu}$ & $\lbrace$ $\beta_{d}, \gamma_{d}, \alpha_{l}$ $\rbrace$ $\lbrace$ $\alpha_{\nu}, \beta_{\nu}, \alpha_{u}, \gamma_{u}$ $\rbrace$ & \\
\cline{1-4}
\multirow{3}{*}{$\lambda_{QL}^{e3C}$} & $\Delta_{3}$ & $\lbrace$ $\beta_{d}$, $\gamma_{d}$, $-\alpha_{\nu}$, $-\beta_{\nu}$, $-\alpha_{l}$, $\alpha_{u}$, $\beta_{u}$ $\rbrace$ & \multirow{3}{*}{
$\lambda_{be}
\left(
\begin{array}{ccc}
0  & 0 & 0  \\
-\frac{V_{tb}}{V_{ts}} & 0 & 0    \\
1 & 0 & 0 
\end{array}
\right)$
} \\
\cline{2-3}
& $\Delta_{3}^{\mu}$  & $\lbrace$ $\beta_{d}$, $\gamma_{d}$, $\alpha_{\nu}$, $\beta_{\nu}$, $\alpha_{l}$, $\alpha_{u}$, $\beta_{u}$ $\rbrace$ & \\
\cline{2-3}
& $\Delta_{1}^{\mu}$ & $\lbrace$ $\beta_{d}, \gamma_{d}, \alpha_{l}$ $\rbrace$ $\lbrace$ $\alpha_{\nu}, \beta_{\nu}, \alpha_{u}, \beta_{u}$ $\rbrace$ & \\
\cline{2-3}
\hline
\hline
\multirow{3}{*}{$\lambda_{QL}^{e\mu 1A}$}  &  $\Delta_{3}$ &
$\lbrace$ $\beta_{d}$, $\gamma_{d}$, $-\beta_{\nu}$,  $-\gamma_{\nu}$,  $-\alpha_{l}$,  $-\beta_{l}$,  $\beta_{u}$,  $\gamma_{u}$ $\rbrace$ 
& \multirow{3}{*}{
$ \lambda_{b\mu}
\left(
\begin{array}{ccc}
0  & 0 & 0  \\
\frac{V_{ub}}{V_{us}} \frac{U_{21}}{U_{11}} & -\frac{V_{ub}}{V_{us}} & 0 \\
 -\frac{U_{21}}{U_{11}} & 1 & 0 
\end{array}
\right)$
} \\
\cline{2-3}
& $\Delta_{3}^{\mu}$  & $\lbrace$ $\beta_{d}$, $\gamma_{d}$, $\beta_{\nu}$,  $\gamma_{\nu}$,  $\alpha_{l}$,  $\beta_{l}$,  $\beta_{u}$,  $\gamma_{u}$ $\rbrace$ & \\
\cline{2-3}
& $\Delta_{1}^{\mu}$ & $\lbrace$ $\beta_{d}, \gamma_{d}, \alpha_{l}, \beta_{l}$ $\rbrace$ $\lbrace$ $\beta_{\nu}, \gamma_{\nu}, \beta_{u}, \gamma_{u}$ $\rbrace$ & \\
\cline{1-4}
\multirow{3}{*}{$ \lambda_{QL}^{e\mu 1B}$} & $\Delta_{3}$ &
$\lbrace$ $\beta_{d}$, $\gamma_{d}$, $-\beta_{\nu}$,  $-\gamma_{\nu}$,  $-\alpha_{l}$,  $-\beta_{l}$,  $\alpha_{u}$,  $\gamma_{u}$ $\rbrace$ 
& \multirow{3}{*}{
$\lambda_{b\mu}
\left(
\begin{array}{ccc}
0  & 0 & 0  \\
\frac{U_{21}}{U_{11}}\frac{V_{cb}}{V_{cs}}  & -\frac{V_{cb}}{V_{cs}} & 0    \\
 -\frac{U_{21}}{U_{11}} & 1 & 0 
\end{array}
\right)$
} \\
\cline{2-3}
& $\Delta_{3}^{\mu}$  & $\lbrace$ $\beta_{d}$, $\gamma_{d}$, $\beta_{\nu}$,  $\gamma_{\nu}$,  $\alpha_{l}$,  $\beta_{l}$,  $\alpha_{u}$,  $\gamma_{u}$ $\rbrace$  & \\
\cline{2-3}
& $\Delta_{1}^{\mu}$ & $\lbrace$ $\beta_{d}, \gamma_{d}, \alpha_{l}, \beta_{l} $ $\rbrace$ $\lbrace$ $\beta_{\nu}, \gamma_{\nu}, \alpha_{u}, \gamma_{u}$ $\rbrace$ & \\
\cline{2-3}
\hline
\hline
\multirow{3}{*}{$\lambda_{QL}^{e\tau 1A}$} & $\Delta_{3}$ &
$\lbrace$ $\beta_{d}$, $\gamma_{d}$, $-\beta_{\nu}$,  $-\gamma_{\nu}$,  $-\alpha_{l}$,  $-\gamma_{l}$,  $\beta_{u}$,  $\gamma_{u}$ $\rbrace$ 
& \multirow{3}{*}{
$\lambda_{b\tau} \left(
\begin{array}{ccc}
0  & 0 & 0 \\
 \frac{U_{31}}{U_{11}} \frac{V_{ub}}{V_{us}} & 0 & -\frac{V_{ub}}{V_{us}}    \\
 -\frac{U_{31}}{U_{11}} & 0 & 1 
\end{array}
\right)$} \\
\cline{2-3}
& $\Delta_{3}^{\mu}$  & $\lbrace$ $\beta_{d}$, $\gamma_{d}$, $\beta_{\nu}$,  $\gamma_{\nu}$,  $\alpha_{l}$,  $\gamma_{l}$,  $\beta_{u}$,  $\gamma_{u}$ $\rbrace$ & \\
\cline{2-3}
& $\Delta_{1}^{\mu}$ & $\lbrace$ $\beta_{d}, \gamma_{d}, \alpha_{l}, \gamma_{l}$ $\rbrace$ $\lbrace$ $\beta_{\nu}, \gamma_{\nu}, \beta_{u}, \gamma_{u}$ $\rbrace$ & \\
\cline{1-4}
\multirow{3}{*}{$\lambda_{QL}^{e\tau 1B}$} & $\Delta_{3}$ &
$\lbrace$ $\beta_{d}$, $\gamma_{d}$, $-\beta_{\nu}$,  $-\gamma_{\nu}$,  $-\alpha_{l}$,  $-\gamma_{l}$,  $\alpha_{u}$,  $\gamma_{u}$ $\rbrace$ 
& \multirow{3}{*}{
$\lambda_{b\tau} \left(
\begin{array}{ccc}
0  & 0 & 0 \\
 \frac{U_{31}}{U_{11}} \frac{V_{cb}}{V_{cs}} & 0 & -\frac{V_{cb}}{V_{cs}}    \\
  -\frac{U_{31}}{U_{11}} & 0 & 1 
\end{array}
\right)$
} \\
\cline{2-3}
& $\Delta_{3}^{\mu}$  & $\lbrace$ $\beta_{d}$, $\gamma_{d}$, $\beta_{\nu}$,  $\gamma_{\nu}$,  $\alpha_{l}$,  $\gamma_{l}$,  $\alpha_{u}$,  $\gamma_{u}$ $\rbrace$ & \\
\cline{2-3}
& $\Delta_{1}^{\mu}$ & $\lbrace$ $\beta_{d}, \gamma_{d}, \alpha_{l}, \gamma_{l}$ $\rbrace$ $\lbrace$ $\beta_{\nu}, \gamma_{\nu}, \alpha_{u}, \gamma_{u}$ $\rbrace$ & \\
\hline
\hline
\multirow{3}{*}{$\lambda_{QL}^{\mu\tau 1A}$} & $\Delta_{3}$ &
$\lbrace$ $\beta_{d}$, $\gamma_{d}$, $-\beta_{\nu}$,  $-\gamma_{\nu}$,  $-\beta_{l}$,  $-\gamma_{l}$,  $\beta_{u}$,  $\gamma_{u}$ $\rbrace$ 
& \multirow{3}{*}{
$\lambda_{b\tau} \left(
\begin{array}{ccc}
0 & 0  & 0  \\
0 &  \frac{U_{31}}{U_{21}} \frac{V_{ub}}{V_{us}} & -\frac{V_{ub}}{V_{us}}    \\
0 & -\frac{U_{31}}{U_{21}} & 1 
\end{array}
\right)$
} \\
\cline{2-3}
& $\Delta_{3}^{\mu}$  & $\lbrace$ $\beta_{d}$, $\gamma_{d}$, $\beta_{\nu}$,  $\gamma_{\nu}$,  $\beta_{l}$,  $\gamma_{l}$,  $\beta_{u}$,  $\gamma_{u}$ $\rbrace$ & \\
\cline{2-3}
& $\Delta_{1}^{\mu}$ & $\lbrace$ $\beta_{d}, \gamma_{d}, \beta_{l}, \gamma_{l}$ $\rbrace$ $\lbrace$ $\beta_{\nu}, \gamma_{\nu}, \beta_{u}, \gamma_{u}$ $\rbrace$ & \\
\cline{1-4}
\multirow{3}{*}{$\lambda_{QL}^{\mu\tau 1B}$} & $\Delta_{3}$ &
$\lbrace$ $\beta_{d}$, $\gamma_{d}$, $-\beta_{\nu}$,  $-\gamma_{\nu}$,  $-\beta_{l}$,  $-\gamma_{l}$,  $\alpha_{u}$,  $\gamma_{u}$ $\rbrace$ 
& \multirow{3}{*}{
$\lambda_{b\tau} \left( 
\begin{array}{ccc}
0 & 0  & 0  \\
0 &  \frac{U_{31}}{U_{21}} \frac{V_{cb}}{V_{cs}} & -\frac{V_{cb}}{V_{cs}}    \\
0 & -\frac{U_{31}}{U_{21}} & 1 
\end{array}
\right)$
} \\
\cline{2-3}
& $\Delta_{3}^{\mu}$  & $\lbrace$ $\beta_{d}$, $\gamma_{d}$, $\beta_{\nu}$,  $\gamma_{\nu}$,  $\beta_{l}$,  $\gamma_{l}$,  $\alpha_{u}$,  $\gamma_{u}$ $\rbrace$ & \\
\cline{2-3}
& $\Delta_{1}^{\mu}$ & $\lbrace$ $\beta_{d}, \gamma_{d}, \beta_{l}, \gamma_{l}$ $\rbrace$ $\lbrace$ $\beta_{\nu}, \gamma_{\nu}, \alpha_{u}, \gamma_{u}$ $\rbrace$ & \\
\cline{2-3}
\hline
\end{tabular}}
\caption{The `fully-reduced' patterns derived in \cite{deMedeirosVarzielas:2019lgb} after the application of SE1 symmetry and experimental constraints, including associated phase equalities required in the generators $T_{a}$ for all leptoquarks considered in this paper. All phases listed in brackets $\lbrace \rbrace$ in the third column are forced to be equal to one another.  NOTE:  $U^{ij}_{PMNS} \equiv U_{ij}$ and $(U^{ij}_{CKM})^{\star} \equiv V_{ij}$.  For the vectors $\Delta_{(1,3)}^{\mu}$, replace $V_{ij} \rightarrow V^{\star}_{ij}$.}
\label{tab:finalphases}
\end{table}
%---------------------------------------------------------------------------------------------------------------------------------------

The matrix elements of \eqref{eq:LQoverconstrain} are of course also constrained by a variety of different experimental observables, in particular lepton flavour violating (LFV) processes (e.g. $\mu \rightarrow e \gamma$), B-meson mixing, and indeed the LNU ratios $\mathcal{R}_{K^{(\star)}}$  --- see \cite{Hiller:2014yaa,Hiller:2017bzc,Varzielas:2015iva,Davidson:1993qk,deMedeirosVarzielas:2019lgb,Hiller:2018wbv} for their specific implications on $\lambda_{dl}$.  Furthermore, when one considers the combined application of \eqref{eq:LQrelations} and \eqref{eq:LQoverconstrain} as is required in SE1, the measured values of the PMNS and CKM matrices become relevant, as the RFS may want to enforce a zero in $\lambda_{QL}$ that cannot be realized experimentally.  All of these considerations have been made in \cite{deMedeirosVarzielas:2019lgb}, where the allowed patterns for $\lambda_{QL}$ were derived in SE1, assuming that they distinguish at least two of three fermion species and that leptoquark couplings mimic SM ones (couplings to heavier fermions are taken to be larger than those to lighter ones).  The explicit matrices obtained in \cite{deMedeirosVarzielas:2019lgb} for $\lambda_{dl}$, as well as all of the associated phase relationships amongst the generators $T_{u,d,l,\nu}$ for the three leptoquarks considered here, are catalogued in Table \ref{tab:finalphases}.  We scan over various NADS that can predict these patterns alongside of special PMNS and CKM matrices in Section \ref{sec:pheno}.  On the other hand, SE2 represents a relaxation of the assumptions made in \cite{deMedeirosVarzielas:2019lgb}.  We will discuss the consequences of this relaxation below and in more detail in Section \ref{sec:relaxscan}, where we also perform another scan to find predictive NADS.  However, both sets of scans described in Section \ref{sec:pheno}-\ref{sec:relaxscan} require us to find a basis where our RFS generators know about the physical mixing patterns we want to connect to $\mathcal{G_{F}}$, precisely as we did above when we rotated to the SM flavour basis in \eqref{eq:SMfermionflav}, so that $T_{\nu U}$ was an explicit function of $U_{PMNS}$.  We now write this basis down.

%%%%%%%%%%%%%%%%%%%%%%%%%%%%%%%%%%%%%%%%%%%%%%%%%%%%%%%%%%%%
\subsection{The Leptoflavour Basis}
\label{sec:LFbasis}
We will in general have new rotations that appear in our leptoquark extension of the SM, namely those that further diagonalize \eqref{eq:genericyuk}. And so, in order to use the reconstruction technique outlined in Section \ref{sec:reconstruction}, we must find a basis where information about these new rotations (and hence about $\lambda_{dl}$) can simultaneously be extracted along with information about the CKM and PMNS matrices of the SM.  

Let us begin in the mass basis of \eqref{eq:LQYUKfull}, where the special patterns of Table \ref{tab:finalphases} were derived, and where each generation of quark and lepton can be uniquely identified.  We recall that here the charged-current interactions of the SM are given by 
\begin{equation}
\mathcal{L}^{CC}_{mass} =  \frac{g}{\sqrt{2}}\bar{l}_L U_{PMNS} \gamma^\mu \nu_L W_{\mu}^- + \frac{g}{\sqrt{2}}\bar{d}_L U_{CKM}^{\dagger} \gamma^\mu u_L W_{\mu}^- + \text{h.c.}
\end{equation}
with the CKM and PMNS matrices defined in \eqref{eq:CKMandPMNS} as the mismatch between up/down and charged lepton/neutrino
mixing matrices, respectively.  In moving to a basis where $\lambda_{dl}$ is generically diagonal, one must be sure to label any further rotations in a manner that respects this (physical) definition.  One way to do so is to rotate fields such that the SM charged currents are simultaneously diagonal with $\lambda_{dl}$, which we refer to as the \emph{leptoflavour basis}.\footnote{This is essentially a basis where all the flavour violation is in the mixing matrices. The physical information is the mismatch between the different sectors and is present in any basis, but the leptoflavour basis is required because the method we employ uses the mixing matrices and thus we need the physical mismatch to be entirely encoded in them.}  This can be achieved by reabsorbing any misalignment introduced in the charged currents by rotations in the charged lepton and down quark sectors via transformations on the neutrino and the up quark fields. 
We therefore construct the leptoflavour basis via the following operations:
\begin{align}
\nonumber
l_L &\rightarrow \Lambda_{l}^{\dagger} l'_L,  &&d_L \rightarrow \Lambda_{d}^\dagger d'_L, &&\nu_L \rightarrow U_{PMNS}^\dagger \Lambda_{l}^\dagger \nu'_L,  &&u_L \rightarrow U_{CKM} \Lambda_{d}^\dagger u'_L, \\
\label{eq:leptoflavourbasistransform}
E_{R} &\rightarrow \Lambda_{E}^{\dagger} E_{R}^{'}, \,\,\,\,\,\,\,\, &&d_R \rightarrow \Lambda_{D}^\dagger d'_R, &&\nu_R \rightarrow  \Lambda_{R}^\dagger \nu'_R, &&u_R \rightarrow  \Lambda_{U}^\dagger u'_R,
\end{align}
where by definition we obtain a new diagonal matrix $\lambda_{dl}^{'}$ given by 
\begin{equation}
\label{eq:dlprime}
\lambda_{dl}^{'} \equiv \Lambda_{d}^* \lambda_{dl} \Lambda_{l}^\dagger.
\end{equation} 
Note that while the RH rotations in \eqref{eq:leptoflavourbasistransform} are not physical in the SM, they can become so in its leptoquark extensions although, for the particular case of the scalar triplet written explicitly below, they are again redundant.  However, this is not the case for the vector singlet, and we therefore include them in all associated equations below for completeness.

Upon applying \eqref{eq:leptoflavourbasistransform}, the corresponding $\Delta_{3}$-enhanced Lagrangian is then found in the leptoflavour basis as 
\begin{align}
\nonumber
\mathcal{L} &\supset  \frac{g}{\sqrt{2}}\bar{l}'_L  \gamma^\mu \nu'_L W_{\mu}^- + \frac{g}{\sqrt{2}}\bar{d}'_L \gamma^\mu u'_L W_{\mu}^- \\
\nonumber
&+ \frac{1}{2}\bar{\nu}_{L}^{'c}\Lambda_{l}^* U_{PMNS}^* m_\nu U_{PMNS}^\dagger \Lambda_{l}^\dagger \nu'_L +\bar{E}^{'}_R \Lambda_{E} m_l \Lambda_{l}^\dagger l'_L + \bar{d}_{R}^{'}\Lambda_{D}  m_d \Lambda_{d}^\dagger d'_L + \bar{u}^{'}_R \Lambda_{U}  m_u U_{CKM}\Lambda_{d}^\dagger u'_L \\
\nonumber
& + \frac{1}{\sqrt{2}}\bar{d}_{L}^{'c} \Lambda_{d}^* \lambda_{dl} \Lambda_{l}^\dagger\nu'_L \Delta_{3}^{1/3} + \bar{d}_{L}^{'c} \Lambda_{d}^* \lambda_{dl} \Lambda_{l}^\dagger l'_L \Delta_{3}^{4/3}  + \bar{u}_{L}^{'c} \Lambda_{d}^* \lambda_{dl} \Lambda_{l}^\dagger\nu'_L \Delta_{3}^{-2/3} + \frac{1}{\sqrt{2}}\bar{u}_{L}^{'c} \Lambda_{d}^* \lambda_{dl} \Lambda_{l}^\dagger l'_L \Delta_{3}^{1/3} \\
\label{eq:Flavour_basis_Lagrangian}
&+\text{h.c.},
\end{align}
where we have already utilized the SU(2) equalities of \eqref{eq:LQrelations}.  We now recall the main assumption of the paper, namely that the SM RFS control (at least one of) the Yukawa-like terms in \eqref{eq:Flavour_basis_Lagrangian} sourced by the leptoquark representation.  In the mass basis this is enforced on the leptoquark terms via \eqref{eq:symmetryexists}, and there is a corresponding relationship in the leptoflavour basis:
\begin{equation}
\label{eq:LQoverconstrainGEN2}
T^{(T,\dagger)'}_{Q}\, \lambda^{'}_{QL}\,T^{'}_{L} \overset{!}{=} \lambda^{'}_{QL} ,
\end{equation} 
with $\lambda^{'}_{QL}$ generically denoting the leptoquark Yukawa couplings in the new basis (c.f. \eqref{eq:dlprime} for the $d-l$ coupling).  The extent to which \eqref{eq:LQoverconstrainGEN2} is explicitly enforced depends on the breaking of $\mathcal{G_{F}}$ to $\mathcal{G}_{a}$ in a complete model, and so we now explore it for the two environments discussed above.

%%%%%%%%%%%%%%%%%%%%%%%%%%%%%%%%%%%%%%%%%%%%%
\subsection*{RFS Invariance in SE1}

In the scenario with fully-reduced matrices, \eqref{eq:LQoverconstrainGEN2} holds $\forall \,\,\,\, \lbrace Q, L \rbrace$, and 
from \eqref{eq:Flavour_basis_Lagrangian} we can then read off the explicit expressions for the leptoflavour basis RFS generators, obtaining
\begin{equation}
\label{Eq:LGenerators_Flavour_basis}
T_{l}^{'} = \Lambda_l T_l \Lambda_{l}^\dagger, \,\,\,\,\,T_{\nu}^{'} = \Lambda_{l} U_{PMNS} T_\nu U_{PMNS}^\dagger \Lambda_{l}^\dagger,\,\,\,\,\, T_{d}^{'}=\Lambda_d T_d \Lambda_{d}^\dagger, \,\,\,\,\, T_{u}^{'} = \Lambda_d U_{CKM}^\dagger T_{u} U_{CKM} \Lambda_{d}^\dagger,
\end{equation}
for the LH generators and 
\begin{equation}
\label{eq:RHgens}
T_{E}^{'}=\Lambda_{E} T_l \Lambda_{E}^{\dagger},\ \ \ \ T_{R}^{'} = \Lambda_{R} T_\nu \Lambda_{R}^{\dagger}, \ \ \ \ T_{D}^{'} = \Lambda_{D} T_d \Lambda_{D}^{\dagger}, \ \ \ \ T_{U}^{'} = \Lambda_{U} T_u \Lambda_{U}^{\dagger},
\end{equation}
for the RH generators ($T_R'$ holds only in the case of Dirac neutrinos). One can easily show that these leave the Lagrangian invariant, as seen explicitly (e.g.) for the $d-\nu$ term:
\begin{align}
\nonumber
 \frac{1}{\sqrt{2}}\bar{d}_{L}^{'c} \Lambda_{d}^* \lambda_{dl} \Lambda_{l}^\dagger\nu'_L \Delta_{3}^{1/3}\,\,\, \longrightarrow \,\,\, & \frac{1}{\sqrt{2}}\bar{d}_{L}^{'c} \Lambda_{d}^* T_{d}^T \Lambda_{d}^T \Lambda_{d}^* \lambda_{dl} \Lambda_{l}^\dagger \Lambda_{l} U_{PMNS} T_\nu U_{PMNS}^\dagger\Lambda_{l}^\dagger\nu'_L \Delta_{3}^{1/3} \\
 \nonumber
 & = \frac{1}{\sqrt{2}}\bar{d}_{L}^{'c} \Lambda_{d}^* T_{d}^T \left[\lambda_{dl} U_{PMNS}\right] T_\nu U_{PMNS}^\dagger\Lambda_{l}^\dagger\nu'_L \Delta_{3}^{1/3} \\
 \nonumber
 & = \frac{1}{\sqrt{2}}\bar{d}_{L}^{'c} \Lambda_{d}^*  \left[\lambda_{dl} U_{PMNS}\right]  U_{PMNS}^\dagger\Lambda_{l}^\dagger\nu'_L \Delta_{3}^{1/3} \\
 \label{eq:dnudemo}
 & = \frac{1}{\sqrt{2}}\bar{d}_{L}^{'c} \Lambda_{d}^* \lambda_{dl} \Lambda_{l}^\dagger\nu'_L \Delta_{3}^{1/3}\,\,\,\,\,\,\,\,\,\,\,\,\text{Q.E.D.}
\end{align}
In moving from the second to third lines we used \eqref{eq:LQrelations} (the bracketed term is simply $\sqrt{2} \lambda_{d\nu}$) and \eqref{eq:LQoverconstrainGEN2}.  Similar equalities hold for all other terms in \eqref{eq:Flavour_basis_Lagrangian}.  We therefore identify \eqref{Eq:LGenerators_Flavour_basis} as the generating set for $\mathcal{G_{F}}$ when RFS are active in all four fermion families, with the phases of $T_{u,d,l,\nu}$ constrained as per Table \ref{tab:finalphases}, and we use them to scan over various possible $\mathcal{G_{F}}$ in Section \ref{sec:pheno} below.  Also note that in the limit where $\Lambda_{d,l} \rightarrow \mathbb{1}$, i.e. the leptoquark couplings are diagonal in the fermion mass basis cf. \eqref{eq:dlprime}, the leptoflavour basis coincides with the flavour basis, and \eqref{Eq:LGenerators_Flavour_basis} returns the generators required to reconstruct a $\mathcal{G_{F}}$ that controls SM mixing only, as expected!  Finally, it is easy to show that the transformations in \eqref{eq:leptoflavourbasistransform} and the resulting generators in \eqref{Eq:LGenerators_Flavour_basis} also hold when considering vector singlet and triplet leptoquarks, since conjugation differences in the corresponding Lagrangians get compensated by the differing SU(2) relationships between couplings, cf. \eqref{eq:LQrelations}-\eqref{eq:LQrelationsV1}.

%%%%%%%%%%%%%%%%%%%%%%%%%%%%%%%%%%%%%%%%%%%%%
\subsection*{RFS Invariance in SE2}
In the scenario with partially-reduced matrices one only demands that  \eqref{eq:LQoverconstrainGEN2} hold for $Q = d$ and $L = l$.  As mentioned above, this can happen when \eqref{eq:LLyukSU2} is enhanced by a single flavon, whose VEV then leaves an overall RFS in $y_{3}^{LL}$ after flavour symmetry breaking.  In this case and upon decomposing isospin indices, moving to the fermion mass basis, and normalizing all couplings to $\lambda_{dl}$, one can easily derive that the RFS acting on the leptoquarks are actioned by:
\begin{equation}
\label{eq:LQgensrelax}
T_{d}^{LQ} = T_{d}, \,\,\,\,\,\,\,\,\,T_{l}^{LQ} = T_{l}, \,\,\,\,\,\,\,\,\,T_{u}^{LQ} = U_{CKM}\,T_{d}^{LQ}\,U_{CKM}^{\dagger},\,\,\,\,\,\,\,\,\,T_{\nu}^{LQ} = U_{PMNS}^{\dagger}\,T_{l}^{LQ}\,U_{PMNS},
\end{equation}
where in general we have been careful to label these operations with `LQ' to distinguish them from the RFS controlling the SM masses, but where in the first two equations we have also already identified the down quark and charged lepton actions with their SM counterparts $T_{d,l}$ (one of our assumptions).  Now, \eqref{eq:Flavour_basis_Lagrangian} of course knows nothing about any RFS, and so the generic shape of the generators in \eqref{Eq:LGenerators_Flavour_basis} also hold in SE2.  However, we must now be careful to distinguish the actions on the SM and leptoquark components of \eqref{eq:Flavour_basis_Lagrangian}.  Plugging \eqref{eq:LQgensrelax} into \eqref{Eq:LGenerators_Flavour_basis} (with appropritae `LQ' labels implied), one immediately sees that the neutrino and up quark generators become redundant:
\begin{equation}
T^{\prime \, LQ}_{\nu} = T^{\prime \, LQ}_{l} =  T^{\prime}_{l}, \,\,\,\,\,\,\,\,\,\,T^{\prime \, LQ}_{u} = T^{\prime\, LQ}_{d} =  T^{\prime}_{d}.
\end{equation}
This is to be expected, since in this symmetry environment we have no way of distinguishing the components of the SU(2) fermion doublets in \eqref{eq:LLyukSU2}.  To see that the invariance of \eqref{eq:Flavour_basis_Lagrangian} still holds under RFS, we repeat the sample calculation above for the $d-\nu$ term:
\begin{align}
\nonumber
 \frac{1}{\sqrt{2}}\bar{d}_{L}^{'c} \Lambda_{d}^* \lambda_{dl} \Lambda_{l}^\dagger\nu'_L \Delta_{3}^{1/3}\,\,\, \longrightarrow \,\,\, & \frac{1}{\sqrt{2}}\bar{d}_{L}^{'c} \Lambda_{d}^* T_{d}^T \Lambda_{d}^T \Lambda_{d}^* \lambda_{dl} \Lambda_{l}^\dagger \Lambda_{l} T_l \Lambda_{l}^\dagger\nu'_L \Delta_{3}^{1/3} \\
 \nonumber
 & = \frac{1}{\sqrt{2}}\bar{d}_{L}^{'c} \Lambda_{d}^* T_{d}^T \left[\lambda_{dl}\right] T_l \Lambda_{l}^\dagger\nu'_L \Delta_{3}^{1/3} \\
 \label{eq:dnudemoB}
 & = \frac{1}{\sqrt{2}}\bar{d}_{L}^{'c} \Lambda_{d}^* \lambda_{dl} \Lambda_{l}^\dagger\nu'_L \Delta_{3}^{1/3}\,\,\,\,\,\,\,\,\,\,\,\,\text{Q.E.D.}
\end{align}
In the second line one notes the subtle difference with respect to \eqref{eq:dnudemo}: the symmetry at work in the $d-\nu$ term is coming from the equality $T_{d}^{T} \lambda_{dl} T_{l} \overset{!}{=} \lambda_{dl}$, not $T_{d}^{T} \lambda_{d\nu} T_{\nu} \overset{!}{=} \lambda_{d\nu}$, which corresponds precisely to the difference in the symmetry assumptions between  SE1 and SE2!  The same is true for the $u-\nu$ and $u-l$ terms not shown, and all invariances again proceed analogously for the vector $\Delta_{(1,3)}^{\mu}$ Lagrangians.\footnote{Note that the distinction between SE1 and SE2 is not meaningful for the RH terms of the vector singlet, as these do not involve SU(2) doublets from the outset.  They are in any event not included in the scans below.}

Of course, the up quark and neutrino mass terms may still be controlled by a respective RFS, and those will still be given by the second and fourth terms in \eqref{Eq:LGenerators_Flavour_basis}.  Therefore, practically speaking, the complete set of generating matrices in the leptoflavour basis are still given by \eqref{Eq:LGenerators_Flavour_basis}-\eqref{eq:RHgens}.  However, there are no longer any phase relationships in $T_{u,d,l,\nu}$ (cf. Table \ref{tab:finalphases}) between any two sectors other than (potentially) the down quarks and charged leptons.  One is also not required to include all four $T_{u,d,l,\nu}^{\prime}$ in the generating set of $\mathcal{G_{F}}$, as it is conceivable that $\mathcal{G_{F}}$ only breaks directly to RFS in certain fermion families.  We will consider three such possibilities in Section \ref{sec:relaxscan}.

%%%%%%%%%%%%%%%%%%%%%%%%%%%%%%%%%%%%%%%%%%%%%
\subsection{On Dirac vs. Majorana Neutrinos}
\label{sec:DiracMass}
While we have chosen to include a Majorana neutrino mass term in the above equations, the analysis proceeds equivalently in the presence of a Dirac mass, whose form is given by
\begin{align}
\nonumber
\mathcal{L} \,\,\, &\supset \,\,\, \bar{\nu}_{R}\,m_{\nu}\,\nu_{L}\,, &&\text{(Fermion mass basis)} \\
\mathcal{L} \,\,\, &\supset \,\,\, \bar{\nu}^{\prime}_{R}\,\Lambda_{R}\,m_{\nu}\,U_{PMNS}^{\dagger}\,\Lambda_{l}^{\dagger}\,\nu^{\prime}_{L}\,,&&\text{(Leptoflavour basis)}
\end{align}  
where we have written it in both the fermion mass and leptoflavour bases. In this case, the neutrinos are Dirac and we do not have the seesaw mechanism at play, we thus require 3 RH neutrinos. Applying \eqref{Eq:LGenerators_Flavour_basis} to the latter, one recovers the original expression as desired:
\begin{align}
\nonumber
 \bar{\nu}^{\prime}_{R}\,\Lambda_{R}\,m_{\nu}\,U_{PMNS}^{\dagger}\,\Lambda_{l}^{\dagger}\,\nu^{\prime}_{L}\,\,\, \longrightarrow \,\,\, &\bar{\nu}^{\prime}_{R}\,\Lambda_{R}\,T_{\nu}^{\dagger}\, \Lambda_{R}^{\dagger}\Lambda_{R}\,m_{\nu}\,U_{PMNS}^{\dagger}\,\Lambda_{l}^{\dagger}\Lambda_{l}\,U_{PMNS}\,T_{\nu}\,U_{PMNS}^{\dagger}\,\Lambda_{l}^{\dagger}\,\nu_{L}^{\prime} \\
 \nonumber
 &= \bar{\nu}^{\prime}_{R}\,\Lambda_{R}\, \left[T_{\nu}^{\dagger}\,m_{\nu}\,T_{\nu}\right] \,U_{PMNS}^{\dagger}\,\Lambda_{l}^{\dagger}\,\nu_{L}^{\prime} \\
 &= \bar{\nu}^{\prime}_{R}\,\Lambda_{R}\,m_{\nu}\,U_{PMNS}^{\dagger}\,\Lambda_{l}^{\dagger}\,\nu_{L}^{\prime}\,\,\,\,\,\,\,\,\,\,\,\,\,\text{Q.E.D.}
\end{align} 
Recall that the equality between the second and third lines is just the natural RFS of the SM masses, cf. \eqref{eq:SMRFS}.  Hence the form of the RFS generators given in \eqref{Eq:LGenerators_Flavour_basis} is the same for both Dirac and Majorana neutrinos.  However, we have already seen in \eqref{eq:SMRFS} that the phases of the fermion mass-basis generators $T_{a}$ potentially differ between the two scenarios, as the maximal RFS for a Majorana mass term is given by a Klein $\mathbb{Z}_{2} \times \mathbb{Z}_{2}$ \cite{Lam:2007qc}.  Indeed, the tacit assumption throughout Sections \ref{sec:massbasis}-\ref{sec:LFbasis} is that $\mathcal{G}_{a}$ is generated by a single matrix representation $T_a$, regardless of whether or not neutrinos are Dirac or Majorana.  In the event it is instead described by a cyclic product group of the form
\begin{equation}
\label{eq:productgroup}
\mathcal{G}_{a} \sim \mathbb{Z}_{a}^{1} \times \mathbb{Z}_{a}^{2} \times \text{...},
\end{equation}
then \eqref{eq:LQoverconstrainGEN2} must be met for each associated $T_{a}^{i\,\prime}$, whose shape is again given by \eqref{Eq:LGenerators_Flavour_basis}, up to the differing phases of the individual $T_{a}^{i}$.  

%%%%%%%%%%%%%%%%%%%%%%%%%%%%%%%%%%%%%%%%%%%%%%%%%%%%%%%%%%%%
\subsection{On Unambiguous Mixing Predictions}
\label{sec:mixamb}
We now wish to emphasize that the complete three-generation fermionic mixing matrices cannot be fully controlled by the RFS of $\mathcal{G_{F}}$ unless all three fermion species are distinguished by the respective $\mathcal{G}_{a}$.  For SM mixing patterns this is perhaps easier to see in the flavour basis \eqref{eq:SMRFS2}, where the generators $T_{aU}$ are functions of the mixing matrices $U_{a}$ predicted. However, if $T_{a}$ has equal phases in its $(i, j)$ entries, then $T_{aU}$ is equivalent to the same matrix rotated through the $(i, j)$ sector:
\begin{equation}
\label{eq:degenEV}
T_{aU} = U_{a}\,T^{ii=jj}_{a} \, U_{a}^{\dagger} = U_{a}\,R_{a}^{ij} \,T^{ii=jj}_{a} \,R_{a}^{ji\star}\, U_{a}^{\dagger}, \,\,\,\,\,\text{with} \,\,\,\,\, R^{ij} \equiv 
\left(
\begin{array}{cc}
\cos \theta_{ij} & \sin \theta_{ij} \,e^{-i\delta_{ij}} \\
-\sin \theta_{ij} \,e^{i\delta_{ij}} & \cos \theta_{ij}
\end{array}
\right). 
\end{equation}
This invariance translates to an ambiguity in the change of basis itself, leading to additional free contributions to the CKM and PMNS matrices. Explicitly, one can write down the transformations to pass from the mass basis to the flavour basis as
\begin{equation}
f_a \rightarrow R_{a}^{ij} U_{a}^\dagger f_{a}^0,
\end{equation}
where $f_{a}^0$ is the usual flavour eigenstate.  One immediately sees that in this case the RFS generator transforms as shown on  the RHS of \eqref{eq:degenEV}, meaning that $\mathcal{G_F}$ cannot unambiguously control fermionic mixing, 
as the predicted CKM and PMNS matrices may still exhibit a dependence on $R_{a}^{ij}$,
\begin{equation}
\label{eq:CKMPMNSamb}
U_{CKM} \Leftrightarrow  R_{u}^{ji\star}\,U_{CKM}\,R_{d}^{mn}, \,\,\,\,\,\,\,\,\,\,\,\,\,\,\,U_{PMNS} \Leftrightarrow  R_{l}^{ji\star}\,U_{PMNS}\,R_{\nu}^{mn}\,,
\end{equation}
that $\mathcal{G_{F}}$ cannot distinguish.  In \eqref{eq:CKMPMNSamb} we are of course not implying that the degeneracies need to be in the same plane for either $T_{u,l}$ nor $T_{d,\nu}$, and clearly $R_{a} = \mathbb{1}$ if $T_{a}$ has three eigenvalues. That is, the RFS controls portions of the mixing, but permits additional free parameter(s). In this case a product group like \eqref{eq:productgroup} would be required for the RFS to pin down an exact $U_{a}$, and in fact this is always true for Majorana neutrinos, since a $\mathbb{Z}_{2}$ symmetry only has two distinct eigenvalues.  Finally, we note that the ambiguity in \eqref{eq:CKMPMNSamb} also holds in the leptoflavour basis that we reconstruct $\mathcal{G_{F}}$ in.

Of course it is entirely plausible that in a complete model the RFS does not control all of the observed mixing, but instead allows free parameters to be fit to data or includes some other mechanism (perhaps auxiliary symmetries) not captured in our simplified framework that solidifies the prediction.  This happens in \cite{Altarelli:2005yx}, for example, where $\mathcal{G_{F}}$ only breaks to $\mathcal{G_{\nu}} \sim \mathbb{Z}_{2}$, but the model unambiguously predicts $U_{PMNS} = U_{TBM}$.  We will therefore state clearly our assumptions in each relevant scan presented in Sections \ref{sec:pheno}-\ref{sec:relaxscan}.

%%%%%%%%%%%%%%%%%%%%%%%%%%%%%%%%%%%%%%%%%%%%%%%%%%%%%%%%%%%%
\section{Closing Finite Groups: the Bottom-Up Approach}
\label{sec:reconstruction}
We now have all relevant information required to close NADS capable of explaining fermionic mixing in the SM and special patterns of leptoquark Yukawa couplings, and to do so we will follow a bottom-up approach that tracks the symmetry breaking backwards in \eqref{eq:GF}, using the generators of $\mathcal{G}_{a}$ to close the larger $\mathcal{G_{F}}$.   We will effectively automate this procedure by taking particular forms for the relevant mixing matrices in question, discretizing the free parameters in those matrices and all phases of $T_{a}$, and scanning over experimentally allowed ranges using the {\tt{GAP}} computational finite algebra package \cite{Sch97,GAP4}.  This is a na\'ive but powerful way to quickly gain information about phenomenologically relevant  $\mathcal{G_{F}}$, and has been applied to matrices in both the lepton \cite{Talbert:2014bda} and quark \cite{Varzielas:2016zuo} sector.  We detail the basic steps below for completeness and to highlight any special points relevant to this new application to leptoquarks.

%%%%%%%%%%%%%%%%%%%%%%%%%%%%%%%%%%%%%%%%%%%%%%%%%%
\subsection{Approximating the CKM and PMNS Matrices \label{sec:CKM_PMNS}} 
\label{sec:mixexperiment}
A key input to \eqref{Eq:LGenerators_Flavour_basis} are the CKM and PMNS mixing matrices of the SM, for which one expects the RFS of $\mathcal{G_{F}}$ to have some control over.  The RFS mechanism was in fact pioneered to search for $\mathcal{G_{F}}$ that can predict their parameters in a model-independent way, and multiple collaborations have used {\tt{GAP}} or other tools/techniques to find such predictive NADS \cite{Lam:2007qc, Ge:2011ih, Ge:2011qn, Hernandez:2012ra,deAdelhartToorop:2011re, Fonseca:2014koa, Hu:2014kca,Lam:2012ga, Holthausen:2012wt, Holthausen:2013vba, King:2013vna, Lavoura:2014kwa, Joshipura:2014pqa, Joshipura:2014qaa, Talbert:2014bda, Yao:2015dwa, King:2016pgv, Varzielas:2016zuo,Yao:2016zev,Lu:2016jit,Li:2017abz,Lu:2018oxc,Hagedorn:2018gpw,Lu:2019gqp}.  The take-away conclusions from those papers are, within the strict (semi-)direct symmetry-breaking approach embodied in \eqref{eq:LQall}-\eqref{eq:LQCKM}, that only large groups of $\mathcal{O}(10^{2})$ are capable of predicting all three measured mixing angles of the PMNS matrix $\theta_{i}^{l}$, while even larger groups are required to explain complete CKM mixing angles $\theta_{i}^{q}$ (or even PMNS mixing simultaneously with the Cabibbo angle).\footnote{Again, flavour models that do not exhibit the symmetry-breaking patterns in \eqref{eq:GF} are not considered in these statements.  Indirect models like that of \cite{deMedeirosVarzielas:2017sdv} can control complete three-generation mixing with small finite groups,  although NLO terms in the OPE still become relevant for the model's phenomenology.}  Hence it may be more natural to consider smaller groups that quantize these matrices to `leading order' (LO), thereby controlling only the dominant observed mixing.  Other smaller mixing angles are then left unconstrained by the RFS, and can either be fitted to free parameters the RFS allows or be realized via other mechanisms that the RFS cannot describe, e.g. Renormalization Group evolution (RGE) from the flavour breaking scale or next-to-leading order (NLO) terms in the operator product expansion (OPE) in flavons defining the effective theory of flavour, which are expected to softly break the RFS.  As regards the former, RGE effects have been well studied,\footnote{Of course, in a complete low-energy phenomenological analysis of leptoquark models one should also consider the RGE of the leptoquark operators/couplings themselves, and not just that of the SM mixing elements.  While it is beyond our scope to do so here, multiple studies (see e.g. \cite{Gonzalez-Alonso:2017iyc,Crivellin:2019kps,Dekens:2018bci,Kumar:2018kmr}) have considered this evolution when applying models to $\mathcal{R}$ observables and/or electric dipole moments.  In \cite{Kumar:2018kmr}, for example, it was found that RGE constraints were in fact less restrictive than LFV constraints in the regime of perturbative couplings for the vector singlet leptoquark.} especially in the context of the SM and its supersymmetric extensions (see e.g. \cite{Olechowski:1990bh,Ross:2007az,Varzielas:2008jm,Chiu:2016qra,Casas:1999tg,Casas:1999ac,Chankowski:1999xc,Antusch:2003kp}), with the conclusion that the magnitude of the quark and lepton sector runnings are highly uncertain due to unknown parameters in both the SM (e.g. the absolute values of the low energy neutrino masses) and SUSY (e.g. the value of $\tan \beta$) --- significant running in the CKM and PMNS matrix elements can be expected from high scales where $\mathcal{G_F}$ is expected to break.  This is especially true in the lepton sector with (nearly-)degenerate neutrino masses. 

Regardless, following the discussion in Section \ref{sec:DiracMass} it is clear from Table \ref{tab:finalphases} and \eqref{Eq:LGenerators_Flavour_basis} that none of the models in SE1 are capable of predicting all three angles in either the CKM or the PMNS matrices anyway; not only do the isolation patterns predict $\theta_{13}^{l} = 0$, but degenerate phases exist in both the quark and lepton sectors, although they are aligned such that the Cabibbo angle of the CKM can (potentially) be predicted in all models except $\lambda_{QL}^{e3C}$.  All two-columned SE1 patterns also permit a free parameter in the (1,3) element of the PMNS matrix which is, when instead \emph{predicted} by the RFS, partially responsible for generating the large (undesirable) groups mentioned above, due to the smallness of the `reactor' angle $\theta_{13}^{l}$.  We will further see in Section \ref{sec:relaxscan} that SE2 environments also require degenerate phases in the quark sector to account for $\mathcal{R}_{K^{(*)}}$.  

It therefore makes sense for us to approximate the forms of the PMNS and CKM matrices in \eqref{Eq:LGenerators_Flavour_basis} in a way that $1)$ is more likely to recover small, natural $\mathcal{G_{F}}$ and $2)$ that can actually capture the unambiguous predictions of most of our simplified models.  To that end we assume the following LO forms:
\begin{align}
\label{eq:mutauPMNS}
U_{PMNS} \simeq U_{\mu \tau} &\equiv \frac{1}{\sqrt{2}}
\left(
\begin{array}{ccc}
\sqrt{2} \cos \theta_{\mu\tau} & \sqrt{2} \sin \theta_{\mu\tau} & 0  \\
-\sin \theta_{\mu\tau} &  \cos \theta_{\mu\tau} & 1   \\
\sin \theta_{\mu\tau}   & - \cos \theta_{\mu\tau}  & 1
\end{array}
\right) + \mathcal{O}\left(\theta_{13}^{l}\right),
\\
\label{eq:Cabibbo}
U_{CKM} \simeq U_{C} &\equiv 
\left(
\begin{array}{ccc}
\cos \theta_{C} & \sin \theta_{C} & 0  \\
- \sin \theta_{C} & \cos \theta_{C} & 0    \\
0   & 0 & 1
\end{array}
\right) + \mathcal{O}\left( \theta_{C}^{2}, \theta_{C}^{3} \right).
\end{align}
The $\mu-\tau$ invariant matrix in \eqref{eq:mutauPMNS} can still provide an excellent description of leptonic mixing up to the small correction required from $\theta_{13}^{l}$.  It includes many popular patterns explored in prior leptonic flavour models, including the tri-bimaximal \cite{Harrison:2002er}, golden ratio \cite{Datta:2003qg,Adulpravitchai:2009bg}, bi-maximal \cite{Fukugita:1998vn}, and hexagonal matrices \cite{Giunti:2002sr,Albright:2010ap}:
\begin{equation}
\label{eq:mt2}
U_{\mu \tau}\left(\theta_{\mu\tau}\right) \rightarrow \begin{cases}
								U_{TBM}  &\rightleftarrows \tan\theta_{\mu\tau} = \frac{1}{\sqrt{2}}
								\\
								U_{BM}  &\rightleftarrows \tan \theta_{\mu\tau}  = 1 \,\,\, \text{or} \,\,\,\theta_{\mu\tau} =\frac{\pi}{4}
								\\
								U_{GR_{1}} &\rightleftarrows \tan\theta_{\mu\tau}= \frac{2}{(1+\sqrt{5})}
								\\
								U_{GR_{2}} &\rightleftarrows \theta_{\mu\tau} =\frac{\pi}{5}
								\\
								U_{HM} &\rightleftarrows \tan\theta_{\mu\tau} = \frac{1}{\sqrt{3}}\,\,\, \text{or} \,\,\,\theta_{\mu\tau} = \frac{\pi}{6}
								\end{cases}
\end{equation}
One observes that any model allowing a free rotation in the (2,3) or (1,3) sectors of this matrix can then successfully account for all experimental constraints on $U_{PMNS}$. 

Similarly, the Cabibbo matrix in \eqref{eq:Cabibbo} describes the dominant CKM mixing between first and second generation quarks excellently, and exterior off-diagonal elements are anyway suppressed by one or two orders of magnitude in comparison.  While free parameter(s) introduced through RFS-allowed rotations of the form in \eqref{eq:degenEV} can further quantize additional element(s), especially in the (2,3) sector, the large hierarchies present in the CKM matrix could also indicate a sub-leading origin for some (or all) of the missing matrix elements in \eqref{eq:Cabibbo}.\footnote{While it may seem strange to simultaneously assume a non-zero $\theta_C$ and zero $\theta_{13}^{l}$, which are similar in magnitude, we recall that the RFS controlling these elements are expected to come from different flavons in different effective operators, and so it is more pertinent to consider the \emph{relative} power suppressions of mixing elements within either the CKM or PMNS matrices.}  

Following on these assumptions we then discretize the free parameters in \eqref{eq:mutauPMNS}-\eqref{eq:Cabibbo} using the schemes in \eqref{eq:mixpi}-\eqref{eq:mixtan}.  Sets of matrices that fulfill the phenomenological constraints we impose, namely
\begin{align}
\label{eq:mutaumixrange}
0.5&\le \sin \theta_{\mu\tau} \le 0.72, \\
\label{eq:Cabmixrange}
0.2 &\le \sin \theta_{C} \le 0.225,
\end{align}
are then collected to form unique mixing matrices, which are then used to form $T_{\nu}^{\prime}$ and $T_{u}^{\prime}$ in \eqref{Eq:LGenerators_Flavour_basis}.  We have chosen a relatively large window for $\sin \theta_{\mu\tau}$ that encompasses all of the L.O. patterns in \eqref{eq:mt2}, and a much narrower window for the (extremely well measured, and typically RGE stable \cite{Olechowski:1990bh,Ross:2007az,Chiu:2016qra}) Cabibbo angle.

%%%%%%%%%%%%%%%%%%%%%%%%%%%%%%%%%%%%%%%%%%%%%%%%%%%%%%%%%%
\subsection{Symmetry Assignment and Discretization}
\label{sec:symassign}
We assign the simplest possible (discrete) RFS to each family sector, namely that mediated by a single cyclic group:
\begin{equation}
\label{eq:cyclic}
\mathcal{G}_{a} \cong \mathbb{Z}_{a}^{n_{a}}
\end{equation}
with $n_{a}$ the order of the symmetry.  Accordingly, the matrices represented by \eqref{Eq:LGenerators_Flavour_basis} are the core group-theoretic and phenomenological engines of our study.

Continuing, we want to find NADS by closing structures generated by the multiple Abelian subgroups of \eqref{eq:cyclic}.  We therefore construct the explicit representations found in \eqref{Eq:LGenerators_Flavour_basis}.  We also intend to exploit the {\tt{SmallGroup}} library of finite groups documented in the {\tt{GAP}}  package, so we must choose a scheme where the free parameters of these matrices (e.g. $\alpha_{d}$, $\beta_{d}$, $...$, $\theta_{\mu\tau}$, $\theta_{C}$, $...$, $\lambda_{s e}/\lambda_{b e}$, $...$) are explicitly quantized, otherwise we would not close finite groups.  Hence we must choose a `discretization scheme' which can be scanned over.  In previous studies \cite{Talbert:2014bda,Varzielas:2016zuo} the generator representations depended only on phases and trigonometric functions (fermionic mixing angles).  For the matrices in \eqref{Eq:LGenerators_Flavour_basis}, however, we must also include the types of parameters entering $\Lambda_{d,l}$, which are just the (generically speaking, unknown) values of ratios of the matrix elements of $\lambda_{dl}$.  We therefore choose the following schemes for the different types of parameters in $T_{a}^{\prime}$, where in all cases we take $\lbrace n,m \rbrace \in Integers$:
\begin{itemize}
\item {\bf{\underline{Leptoquark Matrix Elements}:}}  For the ratios of $\lambda_{dl}$ matrix elements we choose a simple `root-rational' discretization scheme:
\begin{subequations}
\begin{align}
\lambda_{i} &\overset{!}{=} \left(+ \sqrt{\frac{n}{m}}\right)_{i},
\end{align}
\end{subequations}
where the square root operation in {\tt{GAP}} is given by `$ER$' for a rational number, i.e. $\sqrt{n/m} \leftrightarrow ER(n/m)$.  We are therefore implying that these couplings are real, which can be derived as a consequence of SE1 \cite{deMedeirosVarzielas:2019lgb}, but represents a further assumption in SE2.  However, since we have little knowledge of the structure of $\lambda_{dl}$ other than weak bounds on the overall magnitude of some of its elements, this simple scheme will prove sufficient for our current purposes. 
\item {\bf{\underline{Fermionic Mixing Angles}:}} All mixing angles appearing in $U_{CKM}$ and $U_{PMNS}$ are quantized as either
\begin{subequations}
\begin{align}
\label{eq:mixpi}
\theta_{i} &\overset{!}{=} \pi \left(\frac{n}{m}\right)_{i}\,\,\,\text{or} \\
\label{eq:mixtan}
\tan(\theta_{i}) &\overset{!}{=} \left(+\sqrt{\frac{n/m}{1-n/m}} \right)_{i} \, . 
\end{align}
\end{subequations}
In the first scheme we restrict ourselves to $\theta \in \left[ 0, 2\pi \right]$ to avoid degeneracy, and in the second we restrict ourselves to the unit circle.
Of course, these appear in different trigonometric functions in most parameterizations of $U_{CKM}$ and $U_{PMNS}$, so we also give the corresponding {\tt{GAP}} objects for cosines and sines that we construct.  For \eqref{eq:mixpi} one finds
\begin{align}
\nonumber
\cos(n\pi/m) &= \frac{E(2m)^{n}+E(2m)^{-n}}{2}, \\
\sin(n\pi/m) &= \frac{E(2m)^{n}-E(2m)^{-n}}{2E(4)},
\end{align}
with $E(N) = e^{\frac{2\pi i }{N}}$, whereas for \eqref{eq:mixtan} one obtains
\begin{align}
\nonumber
\cos(\theta) &= ER\left(1-\frac{n}{m}\right), \\
\label{eq:mixtanSIN}
\sin(\theta) &= ER\left(\frac{n}{m}\right).
\end{align}
Since in \eqref{eq:mixtan} we restricted ourselves to the unit circle, $n/m \in \left[0,1\right)$ there and in \eqref{eq:mixtanSIN}.
\item {\bf{\underline{Free Phases in RFS Generators}:}} We also quantize the free phases to multiples of $2\pi$ in all fermion mass-basis generators $T_{a}$:
\begin{equation}
\phi_{i} \overset{!}{=} 2\pi \left(\frac{n}{m}\right)_{i} \, .
\end{equation}
Hence we simply create {\tt{GAP}} objects of the form
\begin{equation}
T = \text{diag} \left(E(m)^{n_\alpha}, E(m)^{n_\beta}, E(m)^{n_\gamma} \right)
\end{equation}
in our scripts. 
\end{itemize}
These simple schemes are well-motivated by the representation theory of finite groups, and indeed in Sections \ref{sec:pheno}-\ref{sec:relaxscan} we will show that they are sufficient to reconstruct of a diversity of non-Abelian $\mathcal{G_F}$.  

Given these core parametric inputs, our automation scripts must then have a range of values for $\lbrace n,m \rbrace_{i}$ to scan over.  These domains will not only determine the number of quantizations of $U_{PMNS,CKM}$ and $\Lambda_{d,l}$ entering $T_{a}^{\prime}$, but also even the order $n_{a}$ of the cyclic groups $\mathbb{Z}_{a}$ that get distributed to each family sector.  In all scans in Sections \ref{sec:pheno}-\ref{sec:relaxscan} we choose the following:
\begin{align}
\nonumber
\lbrace n, m \rbrace_{\lambda} &\in \lbrace 1, 1..5 \rbrace \,, \,\,\,\,\,  &&\lbrace n, m \rbrace_{\theta_{C}} \,\,\,\in \lbrace1, 14..15 \rbrace \,, \\
 \label{eq:domains}
\lbrace n, m \rbrace_{\phi_{a}} &\in \lbrace 0..n_{a}, 2.. n_{a} \rbrace \,, \,\,\,\,\, &&\lbrace n, m \rbrace_{\theta_{\mu\tau}} \in \lbrace 1, 1..5 \rbrace \,,
\end{align} 
with $\lambda$ in the first line sometimes called $x$ or $y$ below, and $\phi_a$ representing an arbitrary free phase in a fermion mass-basis generator $T_a$.
While these windows may seem small, they generate a wealth of different group structures, and in any event can be trivially changed given updated experimental or theoretical input.  We then scan across all relevant combinations of \eqref{eq:domains}, and then cull results that do not give phenomenologically relevant quantizations.  This procedure yields a finite number of generating sets $\lbrace T_{a}^{\prime} \rbrace$, where the number of matrices in each set is determined by the symmetry-breaking patterns assumed.

%%%%%%%%%%%%%%%%%%%%%%%%%%%%%%%%%%%%%%%%%%%%%%%%%%
\subsection{Group Closure and Analysis} 
\label{sec:analysis} 
The output of Sections \ref{sec:mixexperiment}-\ref{sec:symassign} are representations for the generators of our RFS that incorporate all relevant symmetry and experimental constraints applicable to the simplified models under consideration.  They are sets of $3 \times 3$ unitary matrices without any free variables --- all have been quantized under one of the above discretization schemes.  Our scripts then collect these unique sets of generators and insist that a parent symmetry $\mathcal{G_{F}}$ is formed from their closure.  To do so we call the $GroupWithGenerators$ command of the {\tt{GAP}} language.  In SE1 we assume the symmetry-breaking patterns in \eqref{eq:GF}, and so the generating set includes four matrices.  On the other hand, in SE2 we are free to assume a variety of different symmetry breaking situations.  For example, it is plausible that the mechanism or symmetry responsible for
 PMNS mixing could have origins independent of that controlling CKM mixing.  For each special pattern of $\lambda_{dl}$ considered, we therefore close the groups generated by the following matrices: 
\begin{align}
\label{eq:LQall}
{\text{({\bf{SE1}}:  Leptoquarks, PMNS, \& CKM):}}\,\,\,\,\,\,\,\,\,\, \mathcal{G}_{F} &\sim \lbrace T_{d}^{\prime}, T_{l}^{\prime},T_{u}^{\prime}, T_{\nu}^{\prime} \rbrace\\
\label{eq:LQallb}
{\text{({\bf{SE1}}:  Leptoquarks, PMNS, \& CKM):}}\,\,\,\,\,\,\,\,\,\, \mathcal{G}_{F} &\sim \lbrace T_{d}^{\prime},T_{u}^{\prime} \rbrace \times \lbrace T_{l}^{\prime}, T_{\nu}^{\prime} \rbrace \\
\label{eq:LQall2}
{\text{({\bf{SE2}}:  Leptoquarks, PMNS, \& CKM):}}\,\,\,\,\,\,\,\,\,\, \mathcal{G}_{F} &\sim \lbrace T_{d}^{\prime}, T_{l}^{\prime},T_{u}^{\prime}, T_{\nu}^{\prime} \rbrace \\
\label{eq:LQall2b}
{\text{({\bf{SE2}}:  Leptoquarks, PMNS, \& CKM):}}\,\,\,\,\,\,\,\,\,\, \mathcal{G}_{F} &\sim \lbrace T_{d}^{\prime}, T_{u}^{\prime} \rbrace \times \lbrace T_{l}^{\prime}, T_{\nu}^{\prime} \rbrace \\
\label{eq:LQPMNS}
{\text{({\bf{SE2}}:  Leptoquarks \& PMNS):}}\,\,\,\,\,\,\,\,\,\, \mathcal{G}_{F} &\sim \lbrace T_{l}^{\prime}, T_{\nu}^{\prime} \rbrace\\
\label{eq:LQCKM}
{\text{({\bf{SE2}}:  Leptoquarks \& CKM):}}\,\,\,\,\,\,\,\,\,\,\mathcal{G}_{F} &\sim \lbrace T_{u}^{\prime}, T_{d}^{\prime} \rbrace
\end{align}
where we have indicated that these closures respectively treat the cases where a single flavour symmetry $\mathcal{G_{F}}$ addresses fermionic mixing and $\mathcal{R}_{K^{(\star)}}$ \eqref{eq:LQall}-\eqref{eq:LQall2b} or either PMNS or CKM mixing alongside of $\mathcal{R}_{K^{(\star)}}$ \eqref{eq:LQPMNS}-\eqref{eq:LQCKM}.  For \eqref{eq:LQall} and \eqref{eq:LQall2} we ask that a single NADS be closed by the generators of all four residual symmetries, whereas in \eqref{eq:LQallb} and \eqref{eq:LQall2b} we consider the case sketched in \eqref{eq:GF}, where $\mathcal{G_{F}} \cong \mathcal{G_{Q}} \times \mathcal{G_{L}}$.  Note that this is not equivalent to simply taking the products of \eqref{eq:LQPMNS} and \eqref{eq:LQCKM}, since additional phase equalities are required amongst $T_a$ when all $\mathcal{G}_a$ are active --- $\mathcal{G_{Q}} \times \mathcal{G_{L}}$ represents a subset of the product of \eqref{eq:LQPMNS} and \eqref{eq:LQCKM}.  In principle we could also define the group $\mathcal{G}_{LQ} \sim \lbrace T_{d}^{\prime}, T_{l}^{\prime} \rbrace$ in SE2, which would have control over $\lambda_{dl}$ and therefore $\mathcal{R}_{K^{(\star)}}$, but no control over fermionic mixing in the SM. However we have found that $\mathcal{G}_{LQ}$ can only be Abelian given our assumptions above and below in Section \ref{sec:relaxscan} --- $T_{d,l}^{\prime}$ are always diagonal --- and so we cannot reconstruct a NADS for $\mathcal{G}_{LQ}$ unless these are softened.

As our CKM parametrization using Cabibbo mixing form only reproduces mixing among the first two generations, we expect that the groups closed using hadronic flavour generators (namely setups \eqref{eq:LQall}-\eqref{eq:LQall2b}) will have 2-dimensional irreducible representations. Groups with 2-dimensional irreducible representations, such as $S_3$ for example, can also lead to predictions for the leptonic sector, for instance $U_{\mu\tau}$ (see Tab.\ \ref{tab:SE2GQLmuon}).

Upon closing the groups in \eqref{eq:LQall}-\eqref{eq:LQCKM} we must still do some culling, as not all will be finite, non-Abelian, of small order, etc.  {\tt{GAP}} includes a number of internal commands that can be used to filter results based on user-defined preferences.   We impose cuts such that we only reconstruct relatively small,\footnote{We note that this will have an impact on the order of the RFS likely to generate the parent group --- allowing larger parent groups will typically yield larger-order $T_a$.  For this reason, we often find $\mathbb{Z}_{2,3,4}$ RFS in the tables of Section \ref{sec:pheno}.}
\begin{equation}
\label{eq:grouporder}
\mathcal{O}\left(\mathcal{G_{F}},\mathcal{G_L} \right) \leq 100,\,\,\, \mathcal{O}\left(\mathcal{G_Q} \right) \leq 50 ,
\end{equation}
and non-Abelian finite groups, and then identify the remaining flavour symmetry candidates with the $GroupID$ and $StructureDescription$ commands.\footnote{Observe that $StructureDescription$ is \emph{not} an isomorphism invariant command; two groups that are not isomorphic can return the same string while isomorphic groups in different representations can return different strings.  %The command is mainly designed to study small groups of $\mathcal{O}(\mathcal{G}_{F}) \lesssim 100$, which is concurrent with our own aims.  
The $GroupID$ command is unique, however.} The latter often returns non-Abelian product structures in terms of Abelian subgroups, and so we recall the corresponding isomorphisms for many common finite group series (see \cite{Ishimori:2010au} for a comprehensive mathematical review of NADS):
\begin{align}
\nonumber
\Sigma(3 N^2) &\cong \left(Z_{N} \times Z_{N} \right) \rtimes Z_{2}, \\
\nonumber
\Delta(3 N^2) &\cong \left(Z_{N} \times Z_{N} \right) \rtimes Z_{3}, \\
\nonumber
\Delta(6 N^2) &\cong \left( \left(Z_{N} \times Z_{N} \right) \rtimes Z_{3} \right) \rtimes Z_{2}, \\
\nonumber
\Sigma(3 N^3) &\cong Z_{N} \times \Delta \left(3 N^2 \right) \,\,\,\text{for}\,\,\,N/3 \neq Integer, \\
\label{eq:groupdefs}
\Sigma(3\cdot3^3) &\cong \left(Z_3 \times Z_3 \times Z_3 \right) \rtimes Z_{3}.
\end{align}  
Note that for brevity we will only report unique combinations of NADS and physical parameter quantizations.  That is, we will not report two results where the same symmetry $\mathcal{G_{F}}$ predicts the same physical parameter(s), but with different phase configurations in the RFS generators $T_{a}$.  Of course these phases are relevant to the additional free parameters that the model allows, cf. \eqref{eq:degenEV}, and so in certain cases we make specific demands about their alignments;  this will be noted when relevant below.  Finally, we also omit results of the form $Z_{N} \times D$, where $D$ is a NADS already identified by the scans. 
  
In addition to giving this information on $\mathcal{G_{F}}$, our scripts also carefully archive the parameters associated to it.  In this way we have all relevant information on the representations of the residual generators, which is necessary if one wishes to construct a consistent model from our results.

%%%%%%%%%%%%%%%%%%%%%%%%%%%%%%%%%%%%%%%%%%%%%%%%%%%%%%%%%%%%
\section{Scanning Fully-Reduced Matrices in SE1}
\label{sec:pheno}

In this section we investigate the different viable leptoquark patterns derived in \cite{deMedeirosVarzielas:2019lgb}. By computing the explicit shape of the leptoquark mixing matrices $\Lambda_{d,l}$, and using the CKM and PMNS assumptions from \eqref{eq:mutauPMNS} and \eqref{eq:Cabibbo}, we  obtain representations for the RFS generators in the leptoflavour basis, which can then be closed to specific group structures as described in Section \ref{sec:analysis}

We obtain $\Lambda_{d,l}$ by utilizing a Singular Value Decomposition (SVD) algorithm, which relies on the fact that a generic matrix $\mathcal{M}$ is diagonalizable by two unitary matrices $\mathcal{U}$ and $\mathcal{V}$,
\begin{equation}
\mathcal{M}^D = \mathcal{U} \mathcal{M} \mathcal{V}^\dagger,
\end{equation}
where $\mathcal{M}^D$ is diagonal. In the event $\mathcal{M}$ is symmetric (or Hermitian, for a $\mathbb{C}$-matrix), only one matrix is required.  In this way we diagonalize the various leptoquark patterns from Table \ref{tab:finalphases} and extract the $\Lambda_d$ and $\Lambda_l$ mixing matrices corresponding to the transformation 
\begin{equation}
\lambda'_{dl} =\Lambda_{d}^* \lambda_{dl} \Lambda_{l}^\dagger,
\end{equation}
where $\lambda'_{dl}$ is diagonal.  We will present the explicit forms of $\Lambda_d$ and $\Lambda_l$ in all cases, before performing the GAP scans.

In what follows we will first study the isolation patterns of Table \ref{tab:finalphases}, and then move on to the two-columned matrices.  We will form groups according to \eqref{eq:LQall}-\eqref{eq:LQallb}.  In all cases we restrict the RFS generators to 
\begin{equation}
\label{eq:RFSorder}
2 \le \mathcal{O}(T_{l,\nu}) \le 5, \,\,\,\,\, 2 \le \mathcal{O}(T_{u,d}) \le 3\,,
\end{equation}
which, when combined with the phenomenological parameter and group-order bounds of Sections \ref{sec:mixexperiment}-\ref{sec:analysis}, yields thousands of generator combinations.  In particular, we scan over 23880(4620) and 42864(8664) RFS generator combinations for $\Delta_3^{(\mu)}$ and $\Delta_1^\mu$, respectively, in the isolation(two-columned) patterns .  In all cases we present our results in tables that include, from left to right, the relevant quantization of the mixing parameters $\theta_{\mu\tau,C}$,\footnote{In this section we only consider the \eqref{eq:mixtan} discretization of $\theta_{\mu\tau}$, which is sufficiently general.  In Section \ref{sec:GLonly} we will study both \eqref{eq:mixpi} and \eqref{eq:mixtan}, observing that the former generates no further groups.} the phase alignments of all four RFS generators $T_{u,d,l,\nu}$, the corresponding GAP {\tt{SmallGroup}} ID of the NADS closed, the common name (GAP \emph{StructureDescription}) for the NADS, and an indication of how many of the $(A,B,C)$ patterns from Table \ref{tab:finalphases} are predicted.

%%%%%%%%%%%%%%%%%%%%%%%%%%%%%%%%%%%%%%%%%%%%%%%%%%%%%%%%%%%
\subsection{Isolation Patterns}
\label{sec:isolation}

%-------------------------------------------------------------------------------------------
\begin{table}[tp]
\renewcommand{\arraystretch}{1.5}
\makebox[\textwidth]{
\begin{tabular}{|c||c|c|c|c||c|c|c|}
\hline
\multicolumn{8}{|c|}{\textbf{Electron Isolation and Fermionic Mixing in SE1}}\\
\hline 
\hline
 $\lbrace t_{\theta_{\mu\tau}}, \theta_{C}  \rbrace$ & $T^{ii}_{l}$ & $T^{ii}_{d}$ &  $T^{ii}_{u}$ &  $T^{ii}_{\nu}$ & GAP-ID & $\mathcal{G_F}$ & A/B \\
\hline 
\hline
$\lbrace\star,\frac{\pi}{14}\rbrace$& [1,1,-1] &[-1,1,1] & [-1,1,1] & [1,1,-1] & [56, 5] & $D_{56}$ & \checkmark/
\checkmark  \\
\hline
\hline
$\lbrace t_{\theta_{\mu\tau}}, \theta_{C}  \rbrace$ & $T^{ii}_{l}$ & $T^{ii}_{d}$ &  $T^{ii}_{u}$ &  $T^{ii}_{\nu}$ & GAP-ID & $\mathcal{G_Q} \times \mathcal{G_L} $ & A/B\\
\hline
\hline
$\lbrace\star,\frac{\pi}{c}\rbrace$& [1,$\omega_3$,$\omega_3^2$] &[-1,1,1] & [-1,1,1] & [1,1,-1] & ([N,d],[6,1]) & $D_{N}\times S_3$ & \checkmark/\checkmark\\
\hline
$\lbrace\star,\frac{\pi}{c}\rbrace$& [1,1,-1] &[-1,1,1] & [-1,1,1] & [1,1,-1] & ([N,d],[8,3]) & $D_{N}\times D_8$ & \checkmark/\checkmark \\
\hline
$\lbrace\star,\frac{\pi}{c}\rbrace$& [1,$\omega_5$,$\omega_5^4$] &[-1,1,1] & [-1,1,1] & [1,1,-1] & ([N,d],[10,1]) & $D_{N}\times D_{10}$ & \checkmark/\checkmark\\
\hline
$\lbrace\star,\frac{\pi}{c}\rbrace$& [1,1,-1] &[-1,1,1] & [-1,1,1] & [1,1,$\omega_4$] & ([N,d],[32,11]) & $D_{N}\times \Sigma(32)$ & \checkmark/\checkmark\\
\hline
$\lbrace\star,\frac{\pi}{c}\rbrace$& [1,$\omega_4$,-$\omega_4$] &[-1,1,1] & [-1,1,1] & [1,1,$\omega_3$] & ([N,d],[36,6]) & $D_{N}\times (Z_3\times(Z_3\rtimes Z_4))$ & \checkmark/\checkmark\\
\hline
$\lbrace\star,\frac{\pi}{c}\rbrace$& [1,1,-1] &[-1,1,1] & [-1,1,1] & [1,1,$\omega_5$] & ([N,d],[50,3]) & $D_{N}\times (Z_5\times D_{10})$ & \checkmark/\checkmark\\
\hline
$\lbrace\star,\frac{\pi}{c}\rbrace$& [1,1,$\omega_4$] &[-1,1,1] & [-1,1,1] & [1,1,$\omega_4$] & ([N,d],[96,67]) & $D_{N}\times (SL^2_3\rtimes Z_4)$ & \checkmark/\checkmark$^\star$\\
\hline
$\lbrace\star,\frac{\pi}{c}\rbrace$& [1,$\omega_4$,-$\omega_4$] &[-1,1,1] & [-1,1,1] & [1,1,$\omega_5$] & ([N,d],[100,6]) & $D_{N}\times (Z_5\times(Z_5\rtimes Z_4))$ & \checkmark/\checkmark \\
\hline
\end{tabular}}
\caption{Flavour symmetries controlling  $\lambda_{dl}^{[e3X]}$, $U_{c}$, and portions of $U_{\mu\tau}$ in SE1. NOTES: $N \in \lbrace 28,30\rbrace$ for all leptoquarks in Pattern A, $N = 14$ for $\Delta_3^{(\mu)}$ in Pattern B, and $N \in \lbrace 14, 28\rbrace$ for $\Delta_1^{\mu}$ in Pattern B.  The corresponding phase alignments are those of $\Delta_3^{(\mu)}$ in Pattern A.  Also, $\lbrace c,d \rbrace =\lbrace N/2 , 3 \rbrace$ for $D_{N=(28,30)}$, and $\lbrace c,d \rbrace =\lbrace N,1  \rbrace$ for $D_{N=14}$.  Finally, the $\checkmark^\star$ notation indicates that the result does not appear for $N=28$, for $\Delta_1^{\mu}$ in Pattern B.}
\label{tab:SE1EISO}
\end{table}
%-------------------------------------------------------------------------------------------

The electron isolation patterns $\lambda_{dl}^{[e3X]}$, with $X=A,B,C$, are given by
\begin{equation}
\lambda_{dl}^{[e3X]}=\lambda_{be}\begin{pmatrix}
0 & 0 & 0 \\
x_X & 0 & 0 \\
1 & 0 & 0 \\
\end{pmatrix},
\ \ \ \mathrm{with} \ \ \ x_X = -\frac{V_{u_X b}}{V_{u_X s}}.
\end{equation}
Performing the SVD decomposition, we find leptoquark mixing matrices of the following forms:
\begin{equation}
\label{eq:SE1Isolambda}
\Lambda_l  = \begin{pmatrix}
1 & 0 & 0 \\
0 & 0 & 1 \\
0 & 1 & 0
\end{pmatrix},
\ \ \
\Lambda_d = \begin{pmatrix}
0 & \frac{x_X}{\sqrt{x_{X}^2+1}} & \frac{1}{\sqrt{x_{X}^2+1}} \\
 0 & -\frac{1}{\sqrt{x_{X}^2+1} \text{sgn}(x_X)} & \frac{1}{\sqrt{1+\frac{1}{x_{X}^2}}} \\
 1 & 0 & 0 
\end{pmatrix}.
\end{equation}
Using our approximations for the CKM and the PMNS matrix one finds that $x_A=x_B=0$ and that $x_C$ is not defined. The mixing matrices then simplify to 
\begin{equation}
\label{eq:LDLLSE1ISO}
\begin{split}
\Lambda_l  = \begin{pmatrix}
1 & 0 & 0 \\
0 & 0 & 1 \\
0 & 1 & 0
\end{pmatrix},
\ \ \ 
\Lambda_d = \begin{pmatrix}
0 & 0 & 1 \\
0 & 1 & 0 \\
1 & 0 & 0
\end{pmatrix}
\end{split}.
\end{equation}
Hence the group scans are only (potentially) sensitive to $\theta_{\mu\tau}$ and $\theta_C$ via $U_{PMNS}$ and $U_{CKM}$.

Forming the leptoflavour RFS generators and closing the groups, one finds the results in Table \ref{tab:SE1EISO}.  As is clear, only $D_{56}$ is closed when a group is formed according to \eqref{eq:LQall}, whereas more diverse structures are permitted when $\mathcal{G_F}\cong \mathcal{G_Q} \times \mathcal{G_F}$, albeit even here only members of the Dihedral series $D_N$ are found for $\mathcal{G_Q}$.  Given that $1)$ $D_N$ groups represent the symmetries of polygons and $2)$ we consider the Cabibbo approximation for $U_{CKM}$, which of course just represents a (discretized) rotation about the angle $\theta_C$ in the (1,2) plane, these results are entirely unsurprising --- see tables below and the results and discussion in \cite{Varzielas:2016zuo,Blum:2007jz}.  The results for $\mathcal{G_L}$ also include Dihedrals, in addition to members of other common finite group series like $S_N$ and $\Sigma(2N^2)$.  More complicated structures are also found, as can be seen in the last four lines of the table.\footnote{Note that $SL^2_3$ is the Special Linear Group of $2\times2$ matrices over the finite field of 3 elements.}

From the phenomenological perspective one observes from the leftmost column that no groups are closed that predict specific values of $\theta_{\mu\tau}$, as indicated by the `$\star$' and as is obvious in the phase alignments of $T_{\nu}^{\prime}$ --- in all cases only the third column of $U_{\mu\tau}$ is controlled by the NADS.  Given the form of \eqref{eq:LDLLSE1ISO} and setting $\beta_\nu = \alpha_\nu$, $T_{\nu}^{\prime}$ is then represented by
\begin{equation}
\label{eq:TnuprimeSE1ISO}
T^{\prime}_{\nu} = 
\left(
\begin{array}{ccc}
e^{i\alpha_{\nu}} & 0 & 0  \\
0 &  \frac{1}{2} \left(e^{i\alpha_\nu} + e^{i\gamma_\nu} \right) & \frac{1}{2}  \left(-e^{i\alpha_\nu} + e^{i\gamma_\nu} \right)    \\
0   & \frac{1}{2}  \left(-e^{i\alpha_\nu} + e^{i\gamma_\nu} \right)   & \frac{1}{2}  \left(e^{i\alpha_\nu} + e^{i\gamma_\nu} \right) 
\end{array}
\right) \, ,
\end{equation}
which is a generalization of the well-known $\mu-\tau$ operator, which clearly knows nothing of $\theta_{\mu\tau}$.  The matrix \eqref{eq:TnuprimeSE1ISO} does however, in the absence of an ambiguity along the lines of \eqref{eq:CKMPMNSamb}, predict $\theta_{13}^{l} = 0$ and $\theta_{23}^{l} = \pi/4$, and this is consistent with the conclusion in \cite{deMedeirosVarzielas:2019lgb} that the RFS of isolation patterns in SE1 predict a null leptonic reactor angle.  On the other hand the groups of Table \ref{tab:SE1EISO} do know about specific quantizations of $\theta_C$, and one observes that even if free parameters exist in both the up and down sectors (assuming there is no other model-specific mechanism that prohibits them), the alignments for $\Delta_3$ in Pattern A (those shown) are such that at least the (1,1) element of $U_c$ is unaffected.  In particular, the $D_N$ groups can predict both values of the Cabibbo angle we allowed for:  $\theta_C \in \pi/14, \pi/15$.  This small set is due to the tight experimental bounds in \eqref{eq:Cabmixrange}. 

%-------------------------------------------------------------------------------------------
\begin{table}[tp]
\renewcommand{\arraystretch}{1.5}
\makebox[\textwidth]{
\begin{tabular}{|c||c|c|c|c||c|c|c|}
\hline
\multicolumn{8}{|c|}{\textbf{$e-\mu$ Patterns and Fermionic Mixing in SE1}}\\
\hline 
\hline
 $\lbrace t_{\theta_{\mu\tau}}, \theta_{C}  \rbrace$ & $T^{ii}_{l}$ & $T^{ii}_{d}$ &  $T^{ii}_{u}$ &  $T^{ii}_{\nu}$ & GAP-ID & $\mathcal{G_{F}}$ & A/B \\
\hline
\hline
  $\lbrace 1, \frac{\pi}{15} \rbrace$ & [1, 1, -1]& [-1, 1, 1]& [-1, 1, 1]& [-1, 1, 1] & [60, 12] & $D_{60}$ & \checkmark/\checkmark \\
\hline
$\lbrace 1, \frac{\pi}{14} \rbrace$ & [1, 1, -1]& [-1, 1, 1]& [-1, 1, 1]& [-1, 1, 1] & [84, 14] & $D_{84}$ & \checkmark/\checkmark \\
\hline
\hline
$\lbrace t_{\theta_{\mu\tau}}, \theta_{C}  \rbrace$ & $T^{ii}_{l}$ & $T^{ii}_{d}$ &  $T^{ii}_{u}$ &  $T^{ii}_{\nu}$ & GAP-ID & $\mathcal{G_Q} \times \mathcal{G_L} $ & A/B\\
\hline
\hline
$\lbrace 1, \frac{\pi}{14} \rbrace$ & [1, 1, -1]& [-1, 1, 1]& [1, -1, 1]& [-1, 1, 1] & ([14,1],[6,1]) & $D_{14}\times S_3$ & \xmark/\checkmark \\
\hline
$\lbrace 1, \frac{\pi}{14} \rbrace$ & [1, 1, -1]& [-1, 1, 1]& [-1, 1, 1]& [-1, 1, 1] & ([28,3],[6,1]) & $D_{28}\times S_3$ & \checkmark/\xmark \\
\hline
$\lbrace 1, \frac{\pi}{15} \rbrace$ & [1, 1, -1]& [-1, 1, 1]& [-1, 1, 1]& [-1, 1, 1] & ([30,3],[6,1]) & $D_{30}\times S_3$ & \checkmark/\xmark \\
\hline
 $\lbrace 1, \frac{\pi}{14} \rbrace$ & [1, 1, -1]& [-1, 1, 1]& [1, -1, -1]& [1, -1, -1] & ([28,3],[12,4]) & $D_{28}\times D_{12}$ & \checkmark$^{\star}$/\checkmark$^{\star}$ \\
\hline
\end{tabular}}
\caption{Flavour symmetries controlling $\lambda_{dl}^{[e\mu X]}$, $U_{C}$, and $U_{\mu \tau}$ in SE1.  When a group is found for both Patterns $A$ and $B$, the phase assignments given are for Pattern $A$.  These results hold for all three leptoquarks, spare the final row, which only appears for $\Delta_1^\mu$ (hence the $\checkmark^\star$).}
\label{tab:SE1emu}
\end{table}
%-------------------------------------------------------------------------------------------

%-------------------------------------------------------------------------------------------
\begin{table}[tp]
\renewcommand{\arraystretch}{1.5}
\makebox[\textwidth]{
\begin{tabular}{|c||c|c|c|c||c|c|c|}
\hline
\multicolumn{8}{|c|}{\textbf{$e-\tau$ Patterns and Fermionic Mixing in SE1 }}\\
\hline 
\hline
 $\lbrace t_{\theta_{\mu\tau}}, \theta_{C}  \rbrace$ & $T^{ii}_{l}$ & $T^{ii}_{d}$ &  $T^{ii}_{u}$ &  $T^{ii}_{\nu}$ & GAP-ID & $\mathcal{G_{F}}$ & A/B \\
\hline
\hline
 $\lbrace 1, \frac{\pi}{15} \rbrace$ & [1, -1, 1]& [-1, 1, 1]& [-1, 1, 1]& [-1, 1, 1] & [30,3] & $D_{30}$ & \parbox[t]{1mm}{\multirow{4}{*}{\rotatebox[origin=c]{90}{See caption.}}}  \\
\cline{1-7}
 $\lbrace 1, \frac{\pi}{14} \rbrace$ & [1, -1, 1]& [-1, 1, 1]& [1, -1, 1]& [-1, 1, 1] & [42,5] & $D_{42}$ & \\
\cline{1-7}
$\lbrace 1, \frac{\pi}{15} \rbrace$ & [1, -1, 1]& [-1, 1, 1]& [1, -1, 1]& [-1, 1, 1] & [60,12] & $D_{60}$ &  \\
\cline{1-7}
 $\lbrace 1, \frac{\pi}{14} \rbrace$ & [1, -1, 1]& [-1, 1, 1]& [-1, 1, 1]& [-1, 1, 1] & [84,14] & $D_{84}$ &  \\
\hline
\hline
$\lbrace t_{\theta_{\mu\tau}}, \theta_{C}  \rbrace$ & $T^{ii}_{l}$ & $T^{ii}_{d}$ &  $T^{ii}_{u}$ &  $T^{ii}_{\nu}$ & GAP-ID & $\mathcal{G_Q} \times \mathcal{G_L} $ & A/B\\
\hline
\hline
 $\lbrace 1, \frac{\pi}{14} \rbrace$ & [1, -1, 1]& [-1, 1, 1]& [1, -1, 1]& [-1, 1, 1] & ([14,1],[6,1]) & $D_{14}\times S_3$ & \xmark/\checkmark \\
\hline
 $\lbrace 1, \frac{\pi}{14} \rbrace$ & [1, -1, 1]& [-1, 1, 1]& [-1, 1, 1]& [-1, 1, 1] & ([28,3],[6,1]) & $D_{28}\times S_3$ & \checkmark/\xmark \\
\hline
 $\lbrace 1, \frac{\pi}{14} \rbrace$ & [1, -1, 1]& [-1, 1, 1]& [1, -1, -1]& [1, -1, -1] & ([28,3],[12,4]) & $D_{28}\times D_{12}$ & \checkmark$^{\star}$/\checkmark$^{\star}$ \\
 \hline
 $\lbrace 1, \frac{\pi}{15} \rbrace$ & [1, -1, 1]& [-1, 1, 1]& [-1, 1, 1]& [-1, 1, 1] & ([30,3],[6,1]) & $D_{30}\times S_3$ & \checkmark/\xmark \\
\hline
\end{tabular}}
\caption{The same as Table \ref{tab:SE1emu}, but for $\lambda_{dl}^{[e\tau X]}$.  For $\mathcal{G_F}$, $D_{30}$ is only found for Pattern A, and $D_{42}$ is only found for Pattern B.  The same is respectively true for $D_{84}$ and $D_{60}$ when considering $\Delta_3^{(\mu)}$, but both are found in both patterns for $\Delta_1^{\mu}$ (we show triplet phases).  The $\checkmark^\star$ notation implies that this group is only found for $\Delta_1^{\mu}$, and the phases correspond to Pattern A.}
\label{tab:SE1etau}
\end{table}
%-------------------------------------------------------------------------------------------

%-------------------------------------------------------------------------------------------
\begin{table}[tp]
\renewcommand{\arraystretch}{1.5}
\makebox[\textwidth]{
\begin{tabular}{|c||c|c|c|c||c|c|c|}
\hline
\multicolumn{8}{|c|}{\textbf{$\mu-\tau$ Patterns and Fermionic Mixing in SE1 }}\\
\hline 
\hline
$\lbrace t_{\theta_{\mu\tau}}, \theta_{C}  \rbrace$ & $T^{ii}_{l}$ & $T^{ii}_{d}$ &  $T^{ii}_{u}$ &  $T^{ii}_{\nu}$ & GAP-ID & $\mathcal{G_{F}}$ & A/B \\
\hline
\hline
 $\lbrace 1, \frac{\pi}{14} \rbrace$ & [1, -1, -1]& [1, -1, -1]& [1, -1, -1]& [1, -1, -1] & [56,5] & $D_{56}$ & \checkmark/\checkmark \\
 \hline
 $\lbrace \frac{1}{\sqrt{3}}, \frac{\pi}{15} \rbrace$ & [1, -1, -1]& [1, -1, -1]& [1, -1, -1]& [1, -1, -1] & [60,12] & $D_{60}$ & \checkmark/\checkmark \\
\hline
 $\lbrace \frac{1}{\sqrt{3}}, \frac{\pi}{14} \rbrace$ & [1, -1, -1]& [1, -1, -1]& [1, -1, -1]& [1, -1, -1] & [84,14] & $D_{84}$ & \checkmark/\checkmark \\
\hline
\hline
 $\lbrace t_{\theta_{\mu\tau}}, \theta_{C}  \rbrace$ & $T^{ii}_{l}$ & $T^{ii}_{d}$ &  $T^{ii}_{u}$ &  $T^{ii}_{\nu}$ & GAP-ID & $\mathcal{G_Q} \times \mathcal{G_L} $ & A/B \\
\hline
\hline
 $\lbrace 1, \frac{\pi}{c} \rbrace$ & [1,-1,-1] & [1,-1,-1] & [-1,1,-1]  & [1,-1,-1]  & ([N,d],[8,3]) & $D_{N} \times D_{8}$ & \parbox[t]{2mm}{\multirow{7}{*}{\rotatebox[origin=c]{90}{See caption below.}}}  \\
\cline{1-7}
 $\lbrace \frac{1}{\sqrt{3}}, \frac{\pi}{c} \rbrace$ & [1,-1,-1] & [1,-1,-1] & [-1,1,-1]  & [1,-1,-1]  & ([N,d],[12,4]) & $D_{N} \times D_{12}$ &  \\
\cline{1-7}
 $\lbrace 1, \frac{\pi}{c} \rbrace$ & [-1,1,1] & [-1,1,1] & [1,-1,1]  & [$\omega_3$,1,1]  & ([N,d],[18,3]) & $D_{N} \times (Z_3\times S_{3})$ & \\
\cline{1-7}
 $\lbrace 1, \frac{\pi}{c} \rbrace$ & [1,-1,-1] & [1,-1,-1] & [-1,1,-1]  & [$\omega_4$,-1,-1]  & ([N,d],[32,11]) & $D_{N} \times \Sigma(32)$ & \\
\cline{1-7}
 $\lbrace 1, \frac{\pi}{c} \rbrace$ & [-1,1,1] & [-1,1,1] & [1,-1,1]  & [$\omega_5$,1,1]  & ([N,d],[50,3]) & $D_{N} \times (Z_5\times D_{10})$ & \\
\cline{1-7}
 $\lbrace \frac{1}{\sqrt{2}}, \frac{\pi}{c} \rbrace$ & [$\omega_3$,1,1] & [-1,1,1] & [1,-1,1] & [$\omega_3$,1,1] & ([N,d],[72,25]) & $D_{N} \times (Z_3 \times SL^2_3)$ &  \\
\cline{1-7}
$\lbrace 1, \frac{\pi}{c} \rbrace$ & [$\omega_4$,1,1] & [-1,1,1] & [1,-1,1]  & [$\omega_4$,1,1]  & ([N,d],[96,67]) & $D_{N} \times (SL_3^2\rtimes Z_{4})$ &  \\
\hline
\end{tabular}}
\caption{The same as Table \ref{tab:SE1emu}, but for $\lambda_{dl}^{[\mu\tau X]}$.  Here $N \in \lbrace 14, 28, 30 \rbrace$, with $N=14$ holding for Pattern B only and $N=28,30$ holding only for Pattern A, except when considering $\Delta_1^\mu$, which also realizes Pattern B when $N=28$, when $\mathcal{G_L}$ is contained in the first five rows of the $\mathcal{G_Q}\times \mathcal{G_L}$ results. $\lbrace c,d \rbrace =\lbrace N/2 , 3 \rbrace$ for $D_{N=(28,30)}$, and $\lbrace c,d \rbrace =\lbrace N,1  \rbrace$ for $D_{N=14}$. The phase alignments in the $\mathcal{G_F}$ section correspond to Pattern A, while those given for $\mathcal{G_Q}$ in the $\mathcal{G_Q}\times \mathcal{G_L}$ section are for $D_{14}$.}
\label{tab:SE1mutau}
\end{table}
%-------------------------------------------------------------------------------------------

%%%%%%%%%%%%%%%%%%%%%%%%%%%%%%%%%%%%%%%%%%%%%%%%%%%%%%%%%%%%
\subsection{Two-Columned Patterns}
\label{sec:SE1twocolumn}

We now investigate the two-columned patterns $\lambda_{dl}^{[ll'1X]}$, where $l,l'= e,\mu,\tau$ and $X=A,B$. There are in total six viable patterns given by
\begin{equation}
\lambda_{dl}^{[e\mu1X]}= \begin{pmatrix}
0 & 0 & 0 \\
x_{X}^{[e\mu]} & y_{X}^{[e\mu]} & 0 \\
z^{[e\mu]} & 1 & 0
\end{pmatrix},
\ \ \ \lambda_{dl}^{[e\tau1X]}= \begin{pmatrix}
0 & 0 & 0 \\
x_{X}^{[e\tau]} & 0 & y_{X}^{[e\tau]}  \\
z^{[e\tau]} & 0 & 1
\end{pmatrix},
\ \ \ \lambda_{dl}^{[\mu\tau1X]}= \begin{pmatrix}
0 & 0 & 0 \\
0 & x_{X}^{[\mu\tau]} & y_{X}^{[\mu\tau]}  \\
0 & z^{[\mu\tau]} & 1
\end{pmatrix},
\end{equation}
where
\begin{equation}
x_{X}^{[l_il_j]} = \frac{V_{u_Xb}}{V_{u_Xs}}\frac{U_{j1}}{U_{i1}}, \ \ \ 
y_{X}^{[l_il_j]} = -\frac{V_{u_Xb}}{V_{u_Xs}}, \ \ \ z^{[l_il_j]} = -\frac{U_{j1}}{U_{i1}}.
\end{equation}
Relying on the CKM and PMNS matrix assumptions in \eqref{eq:mutauPMNS}-\eqref{eq:Cabibbo}, one finds that 
\begin{equation}
x_{X}^{[l_il_j]} = y_{X}^{[l_il_j]} = 0, \ \ \ z^{[e\mu]}=\frac{\tan \theta_{\mu\tau}}{\sqrt{2}}, \ \ \ z^{[e\tau]}=-\frac{\tan \theta_{\mu\tau}}{\sqrt{2}}, \ \ \ z^{[\mu\tau]} = 1.
\end{equation}
Performing the SVD decomposition for each pattern, one obtains
\begin{align}
&\Lambda_l = \begin{pmatrix}
\frac{z}{\sqrt{1+ z^2}} & \frac{1}{\sqrt{1+z^2}}  & 0 \\
 0 & 0 & 1 \\
 -\frac{1}{z\sqrt{1+1/z^2}} & \frac{1}{\sqrt{1+1/z^2}} & 0 \\
\end{pmatrix},
\ \ \
&&\Lambda_d = \begin{pmatrix}
0 & 0 & 1 \\
 0 & 1 & 0 \\
 1 & 0 & 0  \\
\end{pmatrix},\\
&\Lambda_l = \begin{pmatrix}
\frac{z}{\sqrt{1+ z^2}} & 0 & \frac{1}{\sqrt{1+z^2}} \\
 -\frac{1}{z\sqrt{1+1/z^2}} & 0 &  \frac{1}{\sqrt{1+1/z^2}} \\
 0 & 1 & 0 \\
\end{pmatrix},
\ \ \
&&\Lambda_d = \begin{pmatrix}
0 & 0 & 1 \\
 0 & 1 & 0 \\
 1 & 0 & 0 \\
\end{pmatrix},\\
&\Lambda_l = \begin{pmatrix}
 0 & \frac{1}{\sqrt{2}} & \frac{1}{\sqrt{2}} \\
 0 & -\frac{1}{\sqrt{2}} & \frac{1}{\sqrt{2}} \\
 1 & 0 & 0 \\
\end{pmatrix},
\ \ \
&&\Lambda_d = \begin{pmatrix}
0 & 0 & 1 \\
 0 & 1 & 0 \\
 1 & 0 & 0  \\
\end{pmatrix}.
\end{align}
We see that, unlike in the isolation pattern case, information about $\theta_{\mu\tau}$ is communicated to the NADS via both $U_{PMNS}$ and $\Lambda_l$.

Tables \ref{tab:SE1emu}-\ref{tab:SE1mutau} present our results for the $e-\mu$, $e-\tau$, and $\mu-\tau$ patterns, respectively.  We again find that only Dihedral groups are closed when $\mathcal{G_F}$ controls both leptons and quarks simultaneously, but now the NADS does know about both $\theta_{\mu\tau}$ and $\theta_C$.  In particular, for the $e-\mu$ and $e-\tau$ patterns we see that $D_N$ can control bi-maximal $U_{\mu\tau}$ and predict $\theta_C \in \lbrace \pi/14, \pi/15 \rbrace$.  Hexagonal mixing $U_{\mu\tau}$ is also predicted (alongside of the same Cabibbo matrices) for $\lambda_{dl}^{[\mu\tau X]}$.  Finally, the same phenomenology is realized when $\mathcal{G_F}\cong \mathcal{G_Q} \times \mathcal{G_L}$ for $\lambda_{dl}^{[e\mu X]}$ and $\lambda_{dl}^{[e\tau X]}$, but one notices that tri-bimaximal $U_{\mu\tau}$ is also realizable alongside of  $\lambda_{dl}^{[\mu\tau X]}$, when $\mathcal{G_L} \cong Z_3 \times SL^2_3$.  As with the isolation patterns, $D_N$, $S_3$, $\Sigma(32)$, and complicated product groups all appear as leptonic flavour symmetry candidates.

%%%%%%%%%%%%%%%%%%%%%%%%%%%%%%%%%%%%%%%%%%%%%%%%%%%%%%%%%%%%

%-------------------------------------------------------------------------------------------
\begin{table}[tp]
\renewcommand{\arraystretch}{1.5}
\makebox[\textwidth]{
\begin{tabular}{|c||c|c|c||c|c|}
\hline
\multicolumn{6}{|c|}{\textbf{Electron Isolation and Fermionic Mixing in SE2}}\\
\hline
\hline 
 $\lbrace x_{e}, t_{\theta_{\mu\tau}}, \theta_C  \rbrace$ & $T^{ii}_{l}$ &  $T^{ii}_{u}$ &  $T^{ii}_{\nu}$ & GAP-ID & $\mathcal{G_F}$  \\
\hline 
\hline
$\lbrace \frac{1}{2},1,\star\rbrace$ & [-1,1,-1]& [-1,-1,1]& [1,-1,-1]&  [12,4]  & $D_{12}$\\
\hline
$\lbrace1,\star,\star\rbrace$ & [-1,1,-1]& [-1,-1,1]& [-1,-1,1]&  [24,12]  & $S_{4}$  \\
\hline
$\lbrace1,\star,\star\rbrace$ & [-1,$\omega_4$,$\omega_4$]& [-1,-1,1]& [-1,-1,1] &  [96,64]  & $\Delta(96)$\\
\hline
\hline 
 $\lbrace x_{e}, t_{\theta_{\mu\tau}},\theta_C  \rbrace$ & $T^{ii}_{l}$  &  $T^{ii}_{u}$ &  $T^{ii}_{\nu}$ & GAP-ID & $\mathcal{G_Q}\times\mathcal{G_L}$ \\
\hline
\hline
$\lbrace\frac{1}{M},1,\frac{\pi}{c}\rbrace$& [-1,1,-1] & [-1,1,-1] & [1,-1,-1] & ([N,d],[6,1]) & $D_{N}\times S_3$ \\
\hline
$\lbrace\frac{1}{M},1,\frac{\pi}{c}\rbrace$& [-1,1,-1]  & [-1,1,-1] & [$\omega_4$,-$\omega_4$,-1] & ([N,d],[24,12]) & $D_{N}\times S_4$ \\
\hline
$\lbrace\frac{1}{M},1,\frac{\pi}{c}\rbrace$& [-1,$\omega_4$,$\omega_4$]  & [-1,1,-1] & [1,-1,-1] & ([N,d],[32,11]) & $D_{N}\times\Sigma(32)$ \\
\hline
$\lbrace\frac{1}{M},1,\frac{\pi}{c}\rbrace$& [-1,$\omega_4$,$\omega_4$]  & [-1,1,-1] & [1,$\omega_4$,-$\omega_4$] & ([N,d],[96,67]) & $D_{N}\times(SL_3^2\rtimes Z_4)$ \\
\hline
\end{tabular}}
\caption{Flavour symmetries controlling $\lambda_{dl}^{[e0]}$, $U_{c}$, and $U_{\mu\tau}$ in SE2.  In all cases $T_d =\text{diag}(1,-1,-1)$, and the filtered results we present here hold for all three leptoquarks. The variables $\lbrace c,d \rbrace =\lbrace N/2 , 3 \rbrace$ for $D_{N=(28,30)}$, and $\lbrace c,d \rbrace =\lbrace N,1  \rbrace$ for $D_{N=14}$.  The phase alignments shown are for $D_{14}$ --- send $T_u^{ii} \rightarrow [1,-1,-1]$ for $D_{28,30}$. $M \in \lbrace1..5\rbrace$.}
\label{tab:SE2GLGQ}
\end{table}
%-------------------------------------------------------------------------------------------

\section{Scanning Partially-Reduced Matrices in SE2}
\label{sec:relaxscan}
The patterns derived in \cite{deMedeirosVarzielas:2019lgb} are appealing due to their simplicity and predictive power.  However, the assumptions embedded in SE1 are strong, and can be relaxed in explicit models of flavour.  Hence in this section we scan over patterns derived in SE2.  In the corresponding subsections below we will explore three symmetry-breaking environments that fall under the SE2 umbrella:  one where both quarks and leptons are controlled by RFS, and two where \emph{either} quarks or leptons are controlled by RFS.  We discuss the allowed matrices for $\lambda_{dl}$ and the corresponding phase constraints on $T_{d,l}$ following from these assumptions in what follows.  As before, we also give the associated mixing matrices $\Lambda_{d,l}$ derived with an SVD technique, before performing the {\tt{GAP}} scans according to \eqref{eq:LQall2}-\eqref{eq:LQCKM}.  We also respect the RFS group order constraint in \eqref{eq:RFSorder}, except for in Section \ref{sec:GLandGQ} where we limit $2\le \mathcal{O}(T_{l,\nu})\le 4$, and our tables of results have the same organization as above. This yields 22680 different combinations of generators getting scanned over in Section \ref{sec:GLandGQ} for each leptoquark we consider (and in both patterns $\lambda_{dl}^{[e0,\mu0]}$), and either 6640 or 9960 combinations in Section \ref{sec:GLonly}, depending on whether we discretize $\theta_{\mu\tau}$ according to \eqref{eq:mixpi} or \eqref{eq:mixtan}, respectively.  For the simplified pattern studied in Section \ref{sec:GQonly} we only scan over 660 generator combinations.

In addition to these restrictions we further impose that, when scanning through \eqref{eq:LQall2b}-\eqref{eq:LQCKM}, the NADS we reconstruct knows about $\theta_{\mu\tau}$ and/or $\theta_C$.  That is, we demand
\begin{equation}
T_{u}^{11} \neq T_u^{22}\,\,\,\,\,\,\,\,\text{and}\,\,\,\,\,\,\,\,T_{\nu}^{11}\neq T_{\nu}^{22}
\end{equation}
when studying $\mathcal{G_F}\cong\mathcal{G_Q} \times\mathcal{G_L}$ in Section \ref{sec:GLandGQ} and $\mathcal{G_{L,Q}}$ in Sections \ref{sec:GLonly}-\ref{sec:GQonly}.  
%%%%%%%%%%%%%%%%%%%%%%%%%%%%%%%%%%%%%%%%%%%%%%%%%%

%-------------------------------------------------------------------------------------------
\begin{table}[tp]
\renewcommand{\arraystretch}{1.5}
\makebox[\textwidth]{
\begin{tabular}{|c||c|c|c||c|c|}
\hline
\multicolumn{6}{|c|}{\textbf{Muon Isolation and Fermionic Mixing in SE2 ($\mathcal{G_F}$ Case)}}\\
\hline
\hline 
 $\lbrace x_{\mu}, t_{\theta_{\mu\tau}}, \theta_C  \rbrace$ & $T^{ii}_{l}$  &  $T^{ii}_{u}$ &  $T^{ii}_{\nu}$ & GAP-ID & $\mathcal{G_F}$  \\
\hline 
\hline
$\lbrace1,\star,\star\rbrace$ & [-1,-1,1]& [-1,-1,1]& [-1,-1,1]&  [8,3]  & $D_{8}$ \\
\hline
$\lbrace1,1,\star\rbrace$ & [1,-1,-1]& [-1,-1,1]& [1,-1,-1]&  [8,3]  & $D_{8}$ \\
\hline
$\lbrace1,\frac{1}{\sqrt{3}},\star\rbrace$ & [1,-1,-1]& [-1,-1,1]& [1,-1,-1]&  [12,4]  & $D_{12}$ \\
\hline
$\lbrace1,1,\star\rbrace$ & [1,-1,-1]& [-1,-1,1]& [$\omega_3$,$\omega_3^2$,1]&  [12,4]  & $D_{12}$ \\
\hline
$\lbrace1,\frac{1}{\sqrt{3}},\star\rbrace$ & [1,-1,-1]& [-1,-1,1]& [$\omega_4$,-$\omega_4$,1]&  [24,5]  & $Z_4 \times S_{3}$ \\
\hline
$\lbrace\frac{1}{3},\star,\star\rbrace$ & [1,-1,-1]& [-1,-1,1]& [-1,-1,1]&  [24,6]  & $D_{24}$ \\
\hline
$\lbrace1,1,\star\rbrace$ & [-1,-1,1]& [-1,-1,1]& [1,-1,-1]&  [24,12]  & $S_{4}$ \\
\hline
$\lbrace1,\star,\frac{\pi}{14}\rbrace$ & [1,-1,-1]& [1,-1,-1]& [-1,-1,1]&  [28,3]  & $D_{28}$ \\
\hline
$\lbrace1,\frac{1}{\sqrt{3}},\frac{\pi}{15}\rbrace$ & [1,-1,-1]& [1,-1,-1]& [-1,1,-1]&  [30,3]  & $D_{30}$ \\
\hline
$\lbrace1,\star,\star\rbrace$ & [-1,-1,1]& [-1,-1,1]& [$\omega_4$,$\omega_4$,-1]&  [32,11]  & $\Sigma(32)$ \\
\hline
$\lbrace1,1,\star\rbrace$ & [1,-1,-1]& [-1,-1,1]& [1,$\omega_4$,-$\omega_4$]&  [32,11]  & $\Sigma(32)$ \\
\hline
$\lbrace1,1,\star\rbrace$ & [1,-1,-1]& [-1,-1,1]& [1,$\omega_3$,$\omega_3^2$]&  [36,12]  & $Z_6\times S_3$ \\
\hline
$\lbrace1,\frac{1}{\sqrt{3}},\frac{\pi}{14}\rbrace$ & [1,-1,-1]& [-1,1,-1]& [-1,1,-1]&  [42,5]  & $D_{42}$ \\
\hline
$\lbrace1,\star,\frac{\pi}{14}\rbrace$ & [1,-1,-1]& [1,-1,-1]& [$\omega_4$,$\omega_4$,-1]&  [56,4]  & $Z_4 \times D_{14}$ \\
\hline
$\lbrace1,1,\frac{\pi}{14}\rbrace$ & [1,-1,-1]& [1,-1,-1]& [1,-1,-1]&  [56,5]  & $D_{56}$ \\
\hline
$\lbrace1,1,\frac{\pi}{14}\rbrace$ & [1,-1,-1]& [1,-1,-1]& [$\omega_4$,-$\omega_4$,1]&  [56,7]  & $\left(Z_{14} \times Z_2\right)\rtimes Z_2$ \\
\hline
$\lbrace\frac{1}{5},1,\star\rbrace$ & [1,-1,-1]& [-1,-1,1]& [$\omega_3$,$\omega_3^2$,1]&  [60,5]  & $A_{5}$ \\
\hline
$\lbrace\frac{1}{5},\frac{1}{\sqrt{3}},\star\rbrace$ & [1,-1,-1]& [-1,-1,1]& [-1,1,-1]&  [60,5]  & $A_{5}$ \\
\hline
$\lbrace1,\star,\frac{\pi}{15}\rbrace$ & [1,-1,-1]& [1,-1,-1]& [-1,-1,1]&  [60,12]  & $D_{60}$ \\
\hline
$\lbrace1,\frac{1}{\sqrt{3}},\frac{\pi}{15}\rbrace$ & [1,-1,-1]& [1,-1,-1]& [1,-1,-1]&  [60,12]  & $D_{60}$ \\
\hline
$\lbrace1,\frac{1}{\sqrt{3}},\frac{\pi}{14}\rbrace$ & [1,-1,-1]& [1,-1,-1]& [1,-1,-1]&  [84,14]  & $D_{84}$ \\
\hline
$\lbrace1,\star,\star\rbrace$ & [$\omega_4$,-1,$\omega_4$]& [-1,-1,1]& [$\omega_4$,$\omega_4$,-1]&  [96,67]  & $SL^2_3 \rtimes Z_4$ \\
\hline
\end{tabular}}
\caption{The same as in Table \ref{tab:SE2GLGQ} but for the muon isolation pattern. Here we only show reconstructed $\mathcal{G_{F}}$, i.e. those groups formed from the closure of all four RFS generators.}
\label{tab:SE2GQLmuon}
\end{table}
%-------------------------------------------------------------------------------------------

%-------------------------------------------------------------------------------------------
\begin{table}[tp]
\renewcommand{\arraystretch}{1.5}
\makebox[\textwidth]{
\begin{tabular}{|c||c|c|c||c|c|}
\hline
\multicolumn{6}{|c|}{\textbf{Muon Isolation and Fermionic Mixing in SE2 ($\mathcal{G_Q}\times\mathcal{G_L}$ Case)}}\\
\hline
\hline 
 $\lbrace x_{\mu}, t_{\theta_{\mu\tau}}, \theta_C  \rbrace$ & $T^{ii}_{l}$  &  $T^{ii}_{u}$ &  $T^{ii}_{\nu}$ & GAP-ID & $\mathcal{G_Q}\times \mathcal{G_L}$  \\
\hline
\hline
$\lbrace\frac{1}{M},1,\frac{\pi}{c} \rbrace$ & [1,-1,-1] & [-1,1,-1] & [$\omega_3$,$\omega_3^2$,1] & ([N,d],[6,1]) & $D_{N} \times S_3$  \\
\hline
$\lbrace\frac{1}{M},\frac{1}{\sqrt{3}},\frac{\pi}{c} \rbrace$ & [1,-1,-1] & [-1,1,-1] & [-1,1,-1] & ([N,d],[6,1]) & $D_{N} \times S_{3}$  \\
\hline
$\lbrace\frac{1}{M},1,\frac{\pi}{c} \rbrace$ & [1,-1,-1] & [-1,1,-1] & [1,-1,-1] & ([N,d],[8,3]) & $D_{N} \times D_8$  \\
\hline
$\lbrace\frac{1}{M},\frac{1}{\sqrt{3}},\frac{\pi}{c} \rbrace$ & [1,-1,-1] & [-1,1,-1] & [1,-1,-1] & ([N,d],[12,4]) & $D_{N} \times D_{12}$  \\
\hline
$\lbrace\frac{1}{M},1,\frac{\pi}{c} \rbrace$ & [-1,-1,1] & [-1,1,-1] & [$\omega_4$,-$\omega_4$,1] & ([N,d],[24,12]) & $D_{N} \times S_4$  \\
\hline
$\lbrace\frac{1}{M},1,\frac{\pi}{c} \rbrace$ & [1,-1,-1] & [-1,1,-1] & [1,$\omega_4$,-$\omega_4$] & ([N,d],[32,11]) & $D_{N} \times \Sigma(32)$  \\
\hline
\end{tabular}}
\caption{The same as in Table \ref{tab:SE2GQLmuon} but for $\mathcal{G_Q}\times\mathcal{G_L}$ group structures. $\lbrace c,d \rbrace =\lbrace N/2 , 3 \rbrace$ for $D_{N=(28,30)}$, and $\lbrace c,d \rbrace =\lbrace N,1  \rbrace$ for $D_{N=14}$. Again, $M \in \lbrace1..5\rbrace$ and the phase alignments shown are for $D_{14}$ --- send $T_u^{ii} \rightarrow [1,-1,-1]$ for $D_{28,30}$.}
\label{tab:SE2GLGQmuon}
\end{table}
%-------------------------------------------------------------------------------------------

\subsection{Quarks and Leptons}
\label{sec:GLandGQ}
If $\mathcal{G_{F}} \rightarrow \lbrace \mathcal{G}_{u},\mathcal{G}_{d},\mathcal{G}_{l},\mathcal{G}_{\nu}  \rbrace$ we must still satisfy \eqref{eq:LQoverconstrain}, as in SE1.   However, the muon isolation pattern is no longer forbidden and so we obtain
\begin{equation}
\label{eq:yukeisolation0}
\lambda^{[e0]}_{dl} = 
\lambda_{be} \left(
\begin{array}{ccc}
0 & 0 & 0  \\
x_e  & 0 & 0    \\
1  & 0  & 0 
\end{array}
\right), \,\,\,\,\,
\lambda^{[\mu0]}_{dl} = 
 \lambda_{b\mu} \left(
\begin{array}{ccc}
0 & 0 & 0  \\
0 & x_{\mu} & 0    \\
0 & 1 & 0 
\end{array}
\right), \,\,\,\,\, \text{with} \,\,\,\, x_{X} = \frac{\lambda_{sX}}{\lambda_{bX}}.
\end{equation}
These patterns respectively correspond to $-\alpha_{l} = \beta_{d} = \gamma_{d}$ and $-\beta_{l} = \beta_{d} = \gamma_{d}$, for the scalar triplet.  For the vector triplet and singlet the minus signs do not appear in these equalities (as in Table \ref{tab:finalphases}).
However, we are not subject to any further equalities between the phases of $T_{u,\nu}$, and so our overall generating set is not as constrained as in SE1 --- we are still capable of distinguishing three generations of leptons in both the charged and neutrino sectors.  Also note that the quark splitting parameters $x_{X}$ are bound by many experimental constraints --- see the discussion in \cite{Hiller:2014yaa,Varzielas:2015iva,deMedeirosVarzielas:2019lgb}.  In our scans we will demand the following:
\begin{equation}
\label{eq:xybound}
10^{-4} \le x_{X} \le 1\,,
\end{equation}
and consider for the specific scans presented here numbers no smaller than $1/5$ (as seen in Table \ref{tab:SE2GLGQ}), as this generates a sufficient number of interesting groups. Extending or limiting this range is a trivial matter and can be tuned in response to further experimental analysis.

Continuing, we derive the corresponding $\Lambda_{d,l}$ rotations from \eqref{eq:yukeisolation0}, where clearly the matrices in \eqref{eq:SE1Isolambda} hold for the electron isolation pattern $\lambda_{dl}^{[e0]}$ with $x_X \rightarrow x_e$.  For the muon isolation pattern one obtains
\begin{align}
\label{eq:lambdayuke}
&&\Lambda_l = \begin{pmatrix}
0 & 1 & 0 \\
0 & 0 & 1 \\
1 & 0 & 0
\\
\end{pmatrix},
\ \ \
&&&\Lambda_d = \begin{pmatrix}
0 & \frac{x_\mu}{\sqrt{x_{\mu}^2+1}} & \frac{1}{\sqrt{x_{\mu}^2+1}} \\
 0 & -\frac{1}{\sqrt{x_{\mu}^2+1} \text{sgn}(x_\mu)} & \frac{1}{\sqrt{1+\frac{1}{x_{\mu}^2}}} \\
 1 & 0 & 0 
\end{pmatrix}
\end{align}
where as expected only $\Lambda_{l}$ changes from the electron analogue.

%-------------------------------------------------------------------------------------------
\begin{table}[tp]
\renewcommand{\arraystretch}{1.5}
\makebox[\textwidth]{
\begin{tabular}{|c||c|c||c|c|c|}
\hline
\multicolumn{6}{|c|}{\textbf{Lepton Isolation and Lepton Mixing in SE2}}\\
\hline 
\hline
 $\vert\tan\theta_{\mu\tau}\vert$ & $T^{ii}_{l}$  &  $T^{ii}_{\nu}$ & GAP-ID &  $\mathcal{G_{L}}$ & Electron/Muon \\
\hline 
 \hline
  $1/\sqrt{2}$ & $\left[\omega_{3}, 1, \omega_{3}^2 \right]$ & $\left[-1, 1, -1\right]$ & $\left[ 12, 3 \right]$ & $A_{4}$ &  \checkmark /\checkmark \\
  \hline
1  & $\left[1, \omega_{4}, -\omega_{4} \right]$ & $\left[1, -1, -1\right]$ & $\left[ 24, 12 \right]$ & $S_{4}$ & \checkmark /\xmark \\
 \hline
 $1/\sqrt{2}$  & $\left[1, \omega_{3}, \omega_{3}^2 \right]$ & $\left[1, -1, -1\right]$ & $\left[ 24, 12 \right]$ & $S_{4}$ &  \checkmark /\xmark \\
 \hline
 $1/\sqrt{2}$ & $\left[\omega_{3}, 1, \omega_{3}^2\right]$ & $\left[\omega_{4}, -1, \omega_{4}\right]$ & $\left[ 48, 3 \right]$ & $\Delta(48)$ &  \checkmark /\checkmark \\
 \hline
  1  & $\left[\omega_{4}, 1, -1\right]$ & $\left[1, -1, \omega_4\right]$ & $\left[ 48, 30 \right]$ & $A_4 \rtimes Z_{4}$ & \xmark /\checkmark \\
  \hline
    $1/\sqrt{2}$ & $\left[1, \omega_{3}, \omega_{3}^2 \right]$ & $\left[\omega_4, -\omega_4, -\omega_4\right]$ & $\left[ 48, 30 \right]$ & $A_{4} \rtimes Z_4$ &  \checkmark /\xmark \\
 \hline
   $1/\sqrt{2}$  & $\left[\omega_{3}, 1, \omega_{3}^2 \right]$ & $\left[1, -1, -1\right]$ & $\left[ 72, 42 \right]$ & $Z_3 \times S_{4}$ &  \xmark /\checkmark \\
  \hline
    $1/\sqrt{2}$  & $\left[ \omega_{3}, 1, \omega_{3}^2 \right]$ & $\left[\omega_{5}, \omega_{5}^3, \omega_{5}\right]$ & $\left[ 75, 2 \right]$ & $\Delta(75)$ &  \checkmark /\checkmark \\
 \hline
  $1/\sqrt{2}$  & $\left[\omega_{3}, 1, \omega_{3}^2 \right]$ & $\left[1, \omega_{3}, 1\right]$ & $\left[ 81, 7 \right]$ & $\Sigma(81)$ &  \checkmark/\checkmark \\
 \hline
  $1/\sqrt{2}$  & $\left[1, \omega_{3}, \omega_{3}^2 \right]$ & $\left[\omega_{4}, 1, -\omega_{4}\right]$ & $\left[ 96, 64 \right]$ & $\Delta(96)$ &  \checkmark/\xmark\\
 \hline
 1  & $\left[\omega_{4}, 1, -1\right]$ & $\left[1, -1, 1\right]$ & $\left[ 96, 186 \right]$ & $Z_4 \times S_{4}$ & \checkmark /\checkmark \\
 \hline
\end{tabular}}
\caption{Flavour symmetries $\mathcal{G_{L}}$ controlling electron and/or muon isolation patterns $\lambda_{dl}^{[e,\mu]}$ alongside of $U_{\mu \tau}$ lepton mixing in SE2. Note that the phase configurations for $T_{l,\nu}$ are not necessarily equivalent between the electron and muon isolation patterns.  When both are applicable (two \checkmark), we show the phase configurations associated to $\lambda_{dl}^{[\mu]}$.}
\label{tab:GLelectron}
\end{table}
%-------------------------------------------------------------------------------------------

The results of our scans given these inputs are found in Table \ref{tab:SE2GLGQ}-\ref{tab:SE2GLGQmuon}.\footnote{Note that, due to the abundance of viable phase relationships in this symmetry environment, we have further enforced $\text{det}(T_a)=1$ in this Subsection.  This is consistent with the natural expectation that the NADS is a subgroup of a Special Unitary SU(N) group.}  For the electron isolation patterns in Table \ref{tab:SE2GLGQ} one notices that no `four-generator' group $\mathcal{G_F}$ was found that can simultaneously quantize $x_e$, $\theta_{\mu\tau}$, and $\theta_C$.  However, the cubic group $S_4$ and the popular $\Delta(96)$ member of the $\Delta(6N^2)$ series appear for the first time.  These, along with $D_{12}$, can predict the leptoquark coupling ratio $x_e$, and $D_{12}$ can also control bi-maximal mixing.  For $\mathcal{G_F}\cong \mathcal{G_Q}\times \mathcal{G_L}$ one sees that all relevant phenomenological parameters are quantized (as per our assumptions) --- all allowed values of $x_e$ and $\theta_C$ are possible, but only bi-maximal $U_{\mu\tau}$ mixing is found.  In particular, we find that any given $\mathcal{G_Q}\cong D_N$ is capable of controlling any value of $x_e$ at the same value of $\theta_C$, a fact that we have checked explicitly with (non-automated) {\tt{GAP}} scripts and an analytic, `by-hand' closure of $D_{14} \cong \lbrace T_u^\prime, T_d^\prime \rbrace$ at differing $x_e$.  As seen below, similar trends appear for other $\lambda_{dl}$ when considering independent quark symmetries.

The results for muon isolation ($\lambda_{dl}^{[\mu0]}$) in Tables \ref{tab:SE2GQLmuon}-\ref{tab:SE2GLGQmuon} are even richer.  Concentrating on four-generator $\mathcal{G_F}$ in Table \ref{tab:SE2GQLmuon}, we see that Dihedrals are now capable of quantizing all three parameters in our matrices, predicting either bi-maximal or hexagonal $U_{\mu\tau}$ and both $\pi/14$ and $\pi/15$ for $\theta_C$.  

As an illustrative example, we focus on the group $D_{30}$ from Table \ref{tab:SE2GQLmuon}, which makes a prediction for all physical parameters we have isolated.  We use the information regarding $T_a$ given in the table alongside of \eqref{Eq:LGenerators_Flavour_basis} to reconstruct the following $D_{30}$ leptonic generators
\begin{align}
\label{eq:D30leptongens}
T_l^\prime = \begin{pmatrix}
-1 & 0 & 0 \\
0 & -1 & 0 \\
0 & 0 & 1
\\
\end{pmatrix},
\ \
T_\nu^\prime = \frac{1}{2}\begin{pmatrix}
-\frac{1}{2} &-\frac{3}{2} & \sqrt{\frac{3}{2}}  \\
-\frac{3}{2}  & -\frac{1}{2} & -\sqrt{\frac{3}{2}}   \\
  \sqrt{\frac{3}{2}}   &  -\sqrt{\frac{3}{2}}   & -1 
\end{pmatrix},
\end{align}
as well as the quark generators:
\begin{align}
\label{eq:D30quarkgens}
T_d^\prime = \begin{pmatrix}
-1 & 0 & 0 \\
0 & -1 & 0 \\
0 & 0 & 1
\\
\end{pmatrix},
\ \
T_u^\prime = \frac{1}{2}\begin{pmatrix}
-1 - \cos\frac{2\pi}{15} & -1+\cos\frac{2\pi}{15} & \sqrt{2}\sin\frac{2\pi}{15} \\
-1 + \cos\frac{2\pi}{15}  &-1-\cos\frac{2\pi}{15}& -\sqrt{2}\sin\frac{2\pi}{15}  \\
 \sqrt{2}\sin\frac{2\pi}{15}  & -\sqrt{2}\sin\frac{2\pi}{15}    & 2\cos\frac{2\pi}{15}
\end{pmatrix},
\end{align}
of the respective residual symmetries.
These matrices have the specific predictions $x_\mu = 1$, $\theta_{\mu\tau} = \frac{\pi}{6}$, and $\theta_C = \frac{\pi}{15}$ embedded in their representations, as can be seen directly from \eqref{Eq:LGenerators_Flavour_basis}.  One can use these matrices to build a top-down model using $D_{30}$, under the assumption that the vacuum expectation values of the flavour-symmetry breaking flavons respect an invariance under them.

Continuing with Table \ref{tab:SE2GQLmuon}, the product group $\left(Z_{14} \times Z_2 \right) \rtimes Z_2$ also controls the full parameter space.  However, (amongst others) we also notice that the (very small) $D_{8,12}$ groups and the cubic group $S_4$ can predict a unit $x_\mu$ alongside of bi-maximal lepton mixing, and the popular $A_5$ group of the alternating group series $A_N$ appears for the first time, predicting either bi-maximal or hexagonal mixing simultaneously with $x_\mu = 1/5$.  When the group structure is broadened to Table \ref{tab:SE2GLGQmuon} we also see that very small groups in both the quark and lepton sectors yield a rich diversity of phenomenological signatures, including all values of $x_\mu$, as in the electron isolation case.

%%%%%%%%%%%%%%%%%%%%%%%%%%%%%%%%%%%%%%%%%%%%%%%%%%

\subsection{Leptons Only}
\label{sec:GLonly}
When $\mathcal{G_{L}} \rightarrow \lbrace \mathcal{G}_{l}, \mathcal{G}_{\nu} \rbrace $ one can simultaneously control $\lambda_{dl}$ and $U_{PMNS}$. In this case only $T_l$ is active in $\lambda_{dl}$, so satisfying \eqref{eq:LQoverconstrain} is then possible if one or more phases of $T_l$ are set to zero.  Given that we are now only asking $\mathcal{G_L}$ to control portions of SM mixing, we additionally demand that $\mathcal{G}_{l}$ distinguishes all three charged leptons.  This then requires that only \emph{one} phase be set to zero.  Furthermore, satisfying \eqref{eq:LQoverconstrain}  in way that accounts for  $\mathcal{R}_{K^{(*)}}$ means that only $\alpha_{l}$ or $\beta_l$ can be null.  We are therefore led to conclude that only electron and muon isolation patterns are allowed in this environment:
\begin{equation}
\label{eq:yukeisolation}
\lambda^{[e]}_{dl} = 
\lambda_{be}\left(
\begin{array}{ccc}
y_{e} & 0 & 0  \\
x_{e}  & 0 & 0    \\
1 & 0  & 0 
\end{array}
\right), \,\,\,\,\,
\lambda^{[\mu]}_{dl} = 
 \lambda_{b\mu} \left(
\begin{array}{ccc}
0 & y_{\mu} & 0  \\
0 & x_{\mu} & 0    \\
0 & 1 & 0 
\end{array}
\right), \,\,\,\,\, \text{with} \,\,\,\, x_{X} = \frac{\lambda_{sX}}{\lambda_{bX}}, \,\,\, y_{X} = \frac{\lambda_{dX}}{\lambda_{bX}},
\end{equation}
which respectively correspond to $\alpha_l = 0$ and $\beta_l =0$.  Note that, unlike in Section \ref{sec:GLandGQ}, we are no longer forced to set $\lambda_{de}$ or $\lambda_{d\mu}$ to zero, since we have no quark symmetry/phases to differentiate down quarks.   The lack of an active quark symmetry also means that our scan results hold for all leptoquarks under consideration, since the flavour symmetry active in $\lambda_{dl}$ only differentiates between them through relative signs in the down quark and charged lepton generators, cf. Table \ref{tab:finalphases}. It also means we only need to derive $\Lambda_l$ in this scenario, for which we find the pattern in \eqref{eq:SE1Isolambda} holds for $\lambda_{dl}^{[e]}$ and that in \eqref{eq:lambdayuke} for $\lambda_{dl}^{[\mu]}$.  
Note that neither $x_{X}$ nor $y_{X}$ appears in the $\Lambda_{l}$ of \eqref{eq:SE1Isolambda} or \eqref{eq:lambdayuke}, and so the RFS of $\mathcal{G_{L}}$ in this scenario can only control the shape of $\lambda_{dl}$, but not the specific values of its free couplings.

In addition to insisting that $T_{l}$ has three eigenvalues, we will also demand that \emph{either} $1)$ $T_\nu$  has three eigenvalues that can distinguish each neutrino species, and therefore controls a Dirac neutrino mass term with an associated (quantized) $U_{\mu\tau}$ mixing matrix predicted at LO or $2)$ that $T_\nu$ has its phases aligned such that a free parameter can be fitted to $\theta_{13}^l$.  In the latter case we can claim that realistic three-generation PMNS mixing is achievable alongside of controlling $\lambda_{dl}$ at LO.  
 
The results of our {\tt{GAP}} scans are given in Table \ref{tab:GLelectron}, where one observes that a host of NADS have been recovered, including popular groups like $A_4$, $S_4$, $\Sigma(81)$, and more members of the $\Delta(3N^2)$ and $\Delta(6N^2)$ series.  We see that all of the patterns we uncovered are consistent with \eqref{eq:mixtan}, namely the bi-maximal and tri-bimaximal forms of $U_{\mu\tau}$, and we have also given our results for both patterns in \eqref{eq:yukeisolation} in the same table, as many groups were found in common (albeit with slightly different phase configurations).  In particular, we recover the $A_4$ group used in some of the leptoquark models of \cite{Varzielas:2015iva}, including the corresponding VEV alignments.\footnote{These simple $A_4$-based models are again similar to the Altarelli-Feruglio construction \cite{Altarelli:2005yx}, where the quark sector is mostly unadressed with the fields assigned as singlets of $A_4$. The lepton doublet is an $A_4$ triplet and the $A_4$ breaking is communicated differently by distinct $A_4$ triplet flavon VEVs. The extension to leptoquark models in \cite{Varzielas:2015iva} has the same flavon VEV that breaks $A_4$ in the charged lepton sector being used to make the $A_4$ invariant for the terms with the leptoquarks, and this sepcific $A_4$ breaking then leads to lepton isolation patterns for $\lambda_{dl}$.  Specifically, the flavon VEV associated to the charged lepton sector is $ \sim \langle 1, 0, 0 \rangle$, and so the corresponding RFS generator goes as $T_l =\text{diag}\left(1,\omega_3,\omega_3^2 \right)$, which we find for the electron isolation case.}
Of course, lifting some or all of our constraints, in particular the demand for phase alignments in the (1,3) or (2,3) sectors of $T_\nu$, would yield a longer Table \ref{tab:GLelectron}, as would expanding the allowed parameter space for $\theta_{\mu\tau}$ or RFS generator phases our scans populate matrices with.  This latter statement holds for all scans above, as well.

%%%%%%%%%%%%%%%%%%%%%%%%%%%%%%%%%%%%%%%%%%%%%%%%%%

%-------------------------------------------------------------------------------------------
\begin{table}[tp]
\centering
\renewcommand{\arraystretch}{1.5}
\begin{tabular}[t]{|c|c||c|}
\hline
\multicolumn{3}{|c|}{\textbf{$\lambda^{[bs0]}$ and Quark Mixing in SE2}}\\
\hline
\hline
 $\lbrace \theta_{C} \rbrace$ &  $\lbrace y_{b} \rbrace$ &  $\mathcal{G_{Q}} \sim D_{N}$ \\
\hline 
\hline
$ \pi/14 $ & $\lbrace \frac{1}{5}, \frac{1}{4}, \frac{1}{3}, \frac{1}{2}, 1  \rbrace$ & $N \in 14, 28$ \\
\hline
$ \pi/15 $ & $\lbrace \frac{1}{5}, \frac{1}{4}, \frac{1}{3}, \frac{1}{2}, 1  \rbrace$ & $N \in  30$ \\
\hline
\hline
$T^{ii}_{d}$ &  \multicolumn{2}{|c|}{[-1, 1, 1]} \\
\hline
$T^{ii}_{u} \,\,(N=14)$ &  \multicolumn{2}{|c|}{[1, -1, 1]} \\
\hline
$T^{ii}_{u} \,\,(N=28)$ &  \multicolumn{2}{|c|}{[1, -1, -1]} \\
\hline
$T^{ii}_{u} \,\,(N=30)$ &  \multicolumn{2}{|c|}{[-1, 1, 1]} \\
\hline
\end{tabular}\,\,
\,\,\,
\renewcommand{\arraystretch}{1.5}
\begin{tabular}[t]{|c||c|}
\hline
 GAP-ID & $\mathcal{G_{Q}}$ \\
\hline
\hline
  [14, 1] & $D_{14}$ \\
\hline 
 [28, 3] & $D_{28}$ \\
\hline
 [30, 3] & $D_{30}$ \\
\hline
\end{tabular}
\caption{Flavour symmetries $\mathcal{G_{Q}}$ controlling the simplified $\lambda^{[bs0]}$ pattern and $U_{c}$ quark mixing in SE2.}

\label{tab:SE2GQ}
\end{table}
%-------------------------------------------------------------------------------------------

%%%%%%%%%%%%%%%%%%%%%%%%%%%%%%%%%%%%%%%%%%%%%%%%%%%%%%%%%%
\subsection{Quarks Only}
\label{sec:GQonly}
As a final study we consider $\mathcal{G_{Q}} \rightarrow \lbrace \mathcal{G}_{u}, \mathcal{G}_{d} \rbrace$, which can simultaneously control $\lambda_{dl}$ and portions of $U_{CKM}$.  Resolving $\mathcal{R}_{K^{(*)}}$ requires that entries in at least one column of the $s$ and $b$-quark rows be nonzero, and distinguishing two of three quark generations then requires that the all entries of the $d$-quark row be null.  Hence the most general matrix allowed for $\lambda_{dl}$ is given by
\begin{equation}
\label{eq:yukrowisolation}
\lambda^{[bs]}_{dl} = 
 \lambda_{b\tau}  \left(
\begin{array}{ccc}
0 & 0 & 0  \\
x_{s}  & y_{s} & z_{s}    \\
x_{b}  & y_{b} & 1 
\end{array}
\right), \,\,\,\,\, \text{with} \,\,\,\, x_{X} = \frac{\lambda_{Xe}}{\lambda_{b\tau}}, \,\,\, y_{X} = \frac{\lambda_{X\mu}}{\lambda_{b\tau}}, \,\,\, z_{s} = \frac{\lambda_{s\tau}}{\lambda_{b\tau}}.
\end{equation}
This is associated to $\alpha_{d} \neq \beta_{d} = \gamma_{d} = 0$, which in principle permits the determination of the Cabibbo angle, as did all of the simplified models of SE1 except $\lambda_{QL}^{e3C}$.  As in Section \ref{sec:GLonly}, our results hold for all three leptoquarks under consideration. 

The general matrix is hard to work with in an SVD analysis, but we can make a simpler ans\"{a}tz for $\lambda_{dl}^{[bs]}$--- which is motivated by simple flavon models --- as follows:
\begin{equation}
\label{eq:phi23_style}
\lambda^{[bs0]}_{dl} = 
 \lambda_{b\tau}  \left(
\begin{array}{ccc}
0 & 0 & 0  \\
0  & y_{b} & y_{b}  \\
0  & 1 & 1 
\end{array}
\right).
\end{equation}
For the quark splitting parameter we use the bound in \eqref{eq:xybound} with $x_{\mu} \rightarrow y_b$.  The corresponding $\Lambda_d$ is given in \eqref{eq:lambdayuke}, with $x_\mu \rightarrow y_b$.  Unlike in Section \ref{sec:GLonly}, we see that $\mathcal{G_{Q}}$ does have control over the values of the particular leptoquark couplings, and not just the overall shape of $\lambda_{dl}$.

The corresponding $\mathcal{G_{Q}}$ we recover\footnote{We only consider the discretization scheme in \eqref{eq:mixpi}, given prior results in \cite{Varzielas:2016zuo}.} are given in Table \ref{tab:SE2GQ} where we again only find members of the Dihedral series $D_N$, also associated to the two values of the Cabibbo angle we permit:  $\theta_C \in \pi/14, \pi/15$.  However, a number of different quantizations for the quark splitting parameter $y_b$ are found, and so $\mathcal{G_{Q}}$ can easily predict different coupling patterns for $\lambda_{dl}$, and thereby observables like $\mathcal{R}_{K^{(\star)}}$.

%%%%%%%%%%%%%%%%%%%%%%%%%%%%%%%%%%%%%%%%%%%%%%%%%%%%%%%%%%%%
\section{Summary and Outlook}
\label{sec:conclude}
We have shown how the patterns of couplings derived in the `simplified models of flavourful leptoquarks' introduced in \cite{deMedeirosVarzielas:2019lgb} can be sourced from the breakdown of a non-Abelian discrete family symmetry (NADS) $\mathcal{G_{F}}$.  The Abelian residual flavour symmetries (RFS) that remain in the mass terms of SM fermions also control the CKM and PMNS mixing matrices, thereby linking the SM flavour problem with potential observations of lepton non-universality in the $b \rightarrow s ll$ ratio observables $\mathcal{R}_{K^{(*)}}$.  In addition, we have generalized the predictions of \cite{deMedeirosVarzielas:2019lgb} by identifying two classes of simplified models that employ the RFS mechanism:  one where RFS act in all couplings sourced by the original SM-invariant leptoquark terms in \eqref{eq:LLyukSU2}, as in \cite{deMedeirosVarzielas:2019lgb}, and one where the RFS only controls the $\lambda_{dl}$ coupling between down quarks and charged leptons.  We referred to these as Symmetry Environment 1 (SE1) and 2 (SE2) respectively, with the latter representing a highly natural relaxation of the former that can easily be realized in simple flavon-based models.  

Our approach for finding phenomenologically viable NADS follows the strategy outlined in \cite{Talbert:2014bda,Varzielas:2016zuo}, which is automated via scripts written in the {\tt{GAP}} language for computational finite algebra.  Critically, we perform these scans from the bottom-up, meaning that we first specify the subgroup mediating the RFS in different fermion sectors, discretize all available free parameters in a way that respects experimental constraints, and then close parent $\mathcal{G_{F}}$ using the generators of said RFS.  We must do so in a basis where these generators simultaneously know about all predictions we want to connect to $\mathcal{G_{F}}$, and to that end we derived the so-called `leptoflavour' basis where $\lambda_{dl}$ is diagonalized and the physical definitions of the CKM and PMNS matrices are respected.  Our scripts then find a plethora of finite groups that can yield the desired phenomenology upon symmetry breaking, including members of many group series like $D_N$, $A_N$, $S_N$, $\Delta(3N^2)$, $\Delta(6N^2)$, $\Sigma(3N^2)$ and $\Sigma(3N^3)$ that are popular in the flavoured model-building community.  As an important crosscheck, we recover the $A_4$ tetrahedral symmetry and corresponding flavon VEV alignments used in  \cite{Varzielas:2015iva} when we allow for RFS only in the lepton sector, and so our results provide the relevant information necessary to `reconstruct'  complete models of flavour.

However, beyond the imposition of RFS, the approach to studying flavour discussed here and in \cite{deMedeirosVarzielas:2019lgb} is model-independent, as the simplified models we define distill important (falsifiable) phenomenology without committing to additional assumptions regarding the dynamics of flavour-symmetry breaking or any associated UV-complete Lagrangian (which may not be falsifiable). 
Additionally, the ability to structure leptoquark Yukawa couplings, and not just the mixing associated to them, represents a novel and welcome result in comparison to the application of RFS to the SM alone, and may have applications in other BSM constructions (e.g. multi-Higgs-doublet models).  Hence, as the experimental status of LFV, B-meson mixing, and B-decay observables like $\mathcal{R}_{K^{(*)}}$ and the $b \rightarrow cl\nu$ ratio observables $\mathcal{R}_{D^{(*)}}$ \cite{Lees:2012xj,Lees:2013uzd,Huschle:2015rga,Aaij:2015yra,Hirose:2016wfn,Aaij:2017deq,Moriond2019BelleRD} evolve, so will the constraints implied on the various leptoquark couplings, and thereby on the symmetries we employ.   We will leave the exploration of these and other aspects of our simplified models, including their UV-completions and implications at the LHC, to future work.

%%%%%%%%%%%%%%%%%%%%%%%%%%%%%%%%%%%%%%%%%%%%%%%%%%%%%%%%%%%%
\section*{Acknowledgements}
The work of JB is funded by a Ph.D. grant of the French Ministry for Education and Research and supported
 by \textit{Investissements d’avenir}, Labex ENIGMASS, contrat ANR-11-LABX-0012.
IdMV acknowledges
funding from Funda\c{c}\~{a}o para a Ci\^{e}ncia e a Tecnologia (FCT) through the
contract IF/00816/2015 and was supported in part by the National Science Center, Poland, through the HARMONIA project under contract UMO-2015/18/M/ST2/00518 (2016-2019), and by FCT through projects CFTP-FCT Unit 777 (UID/FIS/00777/2019), CERN/FIS-PAR/0004/2017 and PTDC/FIS-PAR/29436/2017 which are partially funded through POCTI (FEDER), COMPETE, QREN and EU.
JT acknowledges research and travel support from DESY, thanks Gino Isidori for interesting insights on the project and its potential extensions, and thanks the CFTP in Lisbon for hospitality and support while portions of this project were completed.
%%%%%%%%%%%%%%%%%%%%%%%%%%%%%%%%%%%%%%%%%%%%%%%%%%%%%%%%%%%%

\end{document}